\documentclass[twocolumn,american,english,aps,prx,superscriptaddress,address,showacknowledgments,longbibliography]{revtex4-1}
\usepackage[T1]{fontenc}
\usepackage[latin9]{inputenc}
\usepackage{color}
\usepackage{amsmath}
\usepackage{amssymb}
\usepackage{graphicx}
\usepackage{esint}

\makeatletter
\@ifundefined{textcolor}{}
{%
 \definecolor{BLACK}{gray}{0}
 \definecolor{WHITE}{gray}{1}
 \definecolor{RED}{rgb}{1,0,0}
 \definecolor{GREEN}{rgb}{0,1,0}
 \definecolor{BLUE}{rgb}{0,0,1}
 \definecolor{CYAN}{cmyk}{1,0,0,0}
 \definecolor{MAGENTA}{cmyk}{0,1,0,0}
 \definecolor{YELLOW}{cmyk}{0,0,1,0}
}

\usepackage{bbm}

\usepackage{babel}

\makeatother

\usepackage{babel}
\begin{document}

\title{Topological Phases of Sound and Light}

\author{V. Peano}

\affiliation{Institute for Theoretical Physics, Friedrich-Alexander-Universität
Erlangen-Nürnberg, Staudtstr. 7, 91058 Erlangen, Germany}

\author{C. Brendel}

\affiliation{Institute for Theoretical Physics, Friedrich-Alexander-Universität
Erlangen-Nürnberg, Staudtstr. 7, 91058 Erlangen, Germany}

\author{M. Schmidt}

\affiliation{Institute for Theoretical Physics, Friedrich-Alexander-Universität
Erlangen-Nürnberg, Staudtstr. 7, 91058 Erlangen, Germany}

\author{F. Marquardt}

\affiliation{Institute for Theoretical Physics, Friedrich-Alexander-Universität
Erlangen-Nürnberg, Staudtstr. 7, 91058 Erlangen, Germany}

\affiliation{Max Planck Institute for the Science of Light, Günther-Scharowsky-Straße
1, 91058 Erlangen, Germany}
\begin{abstract}
Topological states of matter are particularly robust, since they exploit
global features of a material's band structure. Topological states
have already been observed for electrons, atoms, and photons. It is
an outstanding challenge to create a Chern insulator of sound waves
in the solid state. In this work, we propose an implementation based
on cavity optomechanics in a photonic crystal. The topological properties
of the sound waves can be wholly tuned in-situ by adjusting the amplitude
and frequency of a driving laser that controls the optomechanical
interaction between light and sound. The resulting chiral, topologically
protected phonon transport can be probed completely optically. Moreover,
we identify a regime of strong mixing between photon and phonon excitations,
which gives rise to a large set of different topological phases and
offers an example of a Chern insulator produced from the interaction
between two physically distinct particle species, photons and phonons. 
\end{abstract}
\maketitle

\section{Introduction}

Recently, a new paradigm in the classification of the phases of matter
has emerged that is based on topology \cite{Hasan2010RMP}. The Hall
conductance quantization in a 2D electron gas placed inside a magnetic
field is so precise that it serves as a standard to define the Planck
constant. The precision is due to the current being carried by chiral
edge states which are robust against scattering by disorder. It was
realized that at the heart of this effect there is the nontrivial
topology of the bulk electron band structure encoded in topological
invariants, the Chern numbers \cite{Thouless1982}. The modern exploration
of new topological phases started with the prediction of the anomalous
Quantum Hall effect \cite{Haldane1988}. This is a so-called Chern
insulator state that is realized in a staggered magnetic field that
has a vanishing average. The subsequent discovery of the Quantum Spin
Hall effect \cite{Kane2005TopOrder,Bernevig2006} then proved that
even time-reversal symmetry breaking is not necessary. In this case,
the nontrivial topology is induced by the spin-orbit coupling. A third
pathway to a nontrivial topology is the time-dependent modulation
of the band structure in Floquet topological insulators \cite{Oka2009,Gu2011,Lindner2011,Kitagawa2012}.

Inspired by these new developments in our understanding of electronic
systems, researchers have begun to extend the concept to other settings.
Proposals and first experiments on topological phases exist for cold
atoms and ions (e.g. \cite{Bermudez2012,2013_Spielman_Review,2013_Bloch_ZakPhase,2014_Bloch_ChernNumbers,Jotzu2014}).
More closely related to our setting is the theoretical \cite{Haldane2008,Koch2011,Hafezi2011,Umucallar2011,Fang2012,Hafezi2012OptExpr,Khanikaevphotonic2012}
as well as experimental \cite{WangNature2009,Kitagawa2012,Hafezi2013,Rechtsman2013Nat,2014_Lipson_NonreciprocalPhaseShift}
investigation of topologically nontrivial phases of light (see \cite{Lu2014}
for a recent review). Unlike electrons, photons are electrically neutral.
Nevertheless, they mimic the dynamics of charged particles while hopping
on a lattice, e. g. when the time reversal symmetry is broken by synthetic
gauge fields \cite{Haldane2008,Koch2011,Umucallar2011,Fang2012,Hafezi2012OptExpr}
or when an effective spin-orbit coupling is engineered \cite{Hafezi2011,Khanikaevphotonic2012,Hafezi2013}.

At present, it remains an outstanding challenge to engineer topological
phases for sound waves (phonons) in the solid state, with the resulting
robust chiral edge state transport that is useful for applications
in phononics. So far, topological properties have been conjectured
to be present in the vibrations of individual microtubule macromolecules
in biophysics \cite{Prodan2009}, although the precise mechanism requires
further investigation. Moreover, recently it was pointed out that
masses connected by springs or rigid links in special networks (related
to isostatic lattices) show topological features of vibrations. These
include zero-modes localized at some sample edges of an appropriate
geometry \cite{Kane2014}, propagation of topologically protected
nonlinear solitary waves \cite{Chen09092014} in 1D chains, and topologically
robust defect modes bound to dislocations inside a 2D lattice \cite{Paulose2015}.
In contrast to those works, here we are going to propose a 2D phonon
metamaterial of the Chern insulator class which shows chirally propagating
edge states robust against disorder. Very recently, there have been
steps into this direction for macroscopic systems, employing circulating
fluid currents \cite{Yang2015} to break time-reversal invariance,
or wiring up pendula \cite{Huber2015} in the appropriate way to create
a topological insulator. Our goal is to propose a fully tunable nano-scale
system.

It is not trivial to engineer the required non-reciprocal phases for
the transport of phonons in a \foreignlanguage{american}{tunable}
solid-state platform. Although it would be conceivable to employ local
time-dependent modulation of the stress, e.g. using electrodes and
piezoelectric materials (essentially emulating the route towards photonic
magnetic fields proposed in \cite{Fang2012}), this is not very practical,
since the number of wires would scale with the system size.

The tool we are going to employ instead is cavity optomechanics \cite{Aspelmeyer2013RMPArxiv},
a rapidly evolving field that studies the interaction between radiation
and nanomechanical motion, with possible applications in sensing,
classical and quantum communication, and tests of foundational questions
in quantum physics. In particular, we will consider the flexible and
scalable platform of optomechanical crystals \cite{Eichenfield2009,Safavi-Naeini2010APL,Gavartin2011PRL_OMC,2011_Chan_LaserCoolingNanomechOscillator,SafaviNaeini2014SnowCavity}.
These systems are based on free-standing photonic crystals, where
engineered defects support co-localized optical and vibrational modes
interacting via radiation pressure. Recently, it has been proposed
that an array of such point defects would form an optomechanical array
'metamaterial' where the resulting optical and mechanical band structures
could be tuned in-situ by a driving laser \cite{SchmidtarXiv2013}.
Here, we show how to implement a nontrivial topology for sound waves
in a solid state device, based on such optomechanical arrays. This
can be achieved when a suitable lattice geometry is chosen and the
driving laser imprints an appropriate phase pattern on the optomechanical
interaction. The light field then induces an effective Hamiltonian
for the sound waves that leads to a Chern insulator with robust edge
modes. We emphasize that a single laser field (with a suitable phase
pattern) is enough; no time-dependent modulation of any kind is required
in our approach. Our proposal not only presents a practicable route
towards phonon Chern insulators in the solid state, but its realization
would also represent the first example of a topological state of matter
produced using optomechanics.

In addition, we will find that upon sweeping the laser frequency one
can also enter a regime where it is no longer possible to view phonons
and photons as separate. Instead, a whole series of topological phase
transitions arises where both sound and light are involved. This would
be an example of a topologically nontrivial hybrid band structure
made of two physically distinct particle species, with corresponding
edge states for the emerging hybrid excitations. In contrast to the
recently proposed photon-exciton topological polaritons \cite{Karzig2014,Nalitov2015,Bardyn2015},
the interaction in our case is tunable in-situ over a wide range via
the laser amplitude.

\begin{figure*}
\includegraphics[width=1\linewidth]{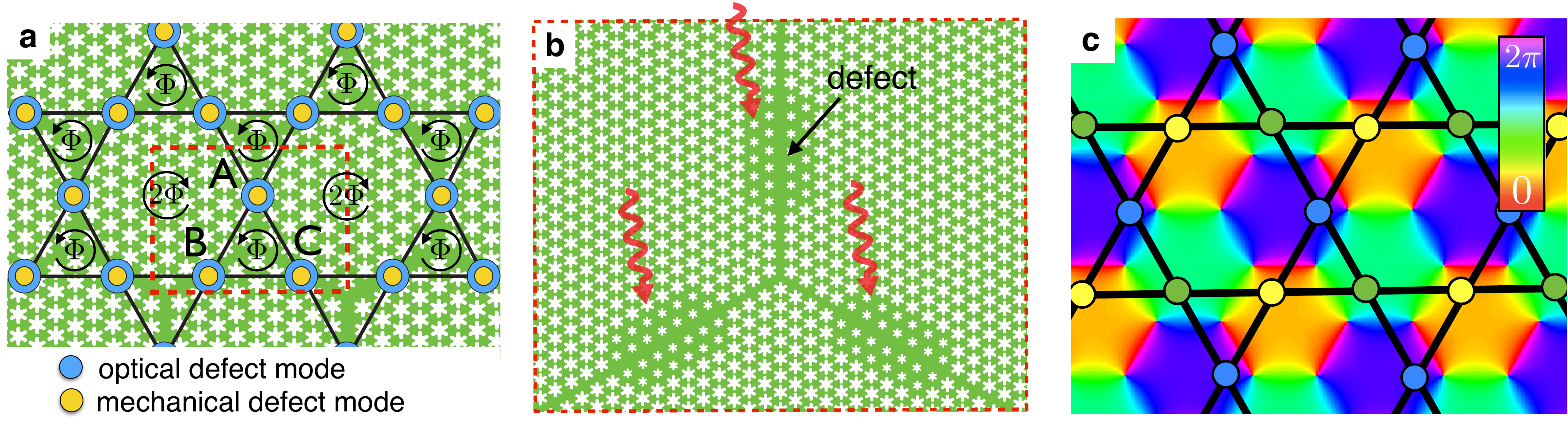}\protect\protect\caption{\label{fig:Setup}A Kagome optomechanical array. (a) Sketch of the
overall arrangement of optical and vibrational modes, with nearest-neighbor
hopping on a Kagome lattice. The effective magnetic fluxes (indicated)
add up to zero, realizing a Chern insulator. (b) Schematic representation
of the elementary building block in a possible realization based on
a 2D snowflake optomechanical crystal (for clarity, the snowflake
size in (a) had been exaggerated in comparison). The picture shows
three linear defects that form at the interfaces between hexagonal
domains of a periodic snowflake hole pattern. In the center of each,
there is an engineered localized point-like defect mode (as in the
experiment of \cite{SafaviNaeini2014SnowCavity}). (c) Suitable optical
phase pattern, generated by the superposition of three beams meeting
at 120 degree angles in the plane of the sample (illustrated here
for a scalar field).}
\end{figure*}

\section{Results}

\subsection{Optomechanical Arrays}

In the field of cavity optomechanics \cite{Aspelmeyer2013RMPArxiv}
the basic interaction between light and mechanical motion comes about
because any deformation of an optical cavity's boundaries will lead
to a shift of the cavity's optical mode frequencies. Focussing on
a single cavity mode, its energy may therefore be expressed as $\hbar\omega_{{\rm cav}}(\hat{x})\hat{a}^{\dagger}\hat{a}$,
where $\hat{x}$ represents the mechanical displacement and $\hat{a}^{\dagger}\hat{a}$
is the photon number. Expanding to leading order in $\hat{x}$, which
is usually an excellent approximation \cite{Aspelmeyer2013RMPArxiv},
this yields the basic interaction $\hbar\omega_{{\rm cav}}'\hat{x}\hat{a}^{\dagger}\hat{a}$.
The mechanical motion is very often dominated by a single harmonic
vibration mode, such that $\hat{x}=x_{{\rm ZPF}}(\hat{b}+\hat{b}^{\dagger})$,
with $x_{{\rm ZPF}}$ the mechanical zero-point fluctuations and $\hat{b}$
the phonon annihilation operator. Thus one arrives at the fundamental
optomechanical interaction 
\begin{equation}
-\hbar g_{0}\hat{a}^{\dagger}\hat{a}(\hat{b}+\hat{b}^{\dagger})\,,\label{eq:StandardOMHamiltonian}
\end{equation}
where $g_{0}=-\omega_{{\rm cav}}'x_{{\rm ZPF}}$ is the bare coupling
constant. The optomechanical coupling rate $g_{0}$ represents the
optical shift due to a mechanical zero-point displacement, and it
is typically much smaller than the photon decay rate $\kappa$. However,
by illuminating the sample with laser light, one can effectively enhance
the optomechanical interaction. When the system is driven by a laser,
one can write $\hat{a}=\alpha+\delta\hat{a}$, where $\alpha$ is
the complex amplitude set by the laser drive and $\delta\hat{a}$
represents the quantum fluctuations on top of that. Keeping the leading
nontrivial terms, one obtains a quadratic Hamiltonian (the so-called
'linearized optomechanical interaction'),

\begin{equation}
-\hbar g_{0}(\alpha^{*}\delta\hat{a}+\alpha\delta\hat{a}^{\dagger})(\hat{b}+\hat{b}^{\dagger})\,.\label{eq:LinearizedInteraction}
\end{equation}
This is the well-tested basis for the description of almost all quantum-optomechanical
experiments to date \cite{Aspelmeyer2013RMPArxiv}. The new, effective
coupling constant $g=g_{0}\alpha$ is laser-tunable and may be complex,
containing a phase factor set by the laser phase, which will become
crucial in our scheme. Eq.~(\ref{eq:LinearizedInteraction}) describes
the interconversion between phonons and photon excitations at the
cavity mode frequency (terms $\delta\hat{a}^{\dagger}\hat{b}$ and
$\delta\hat{a}\hat{b}^{\dagger}$). Physically, these conversion processes
can be understood as anti-Stokes Raman transitions, where the driving
photons impinging on the cavity are inelastically scattered into higher-frequency
photons by absorbing a phonon (enabling e.g. laser-cooling of mechanical
motion). Depending on the laser frequency, there can also be Stokes
processes, where driving photons are scattered to lower frequencies
while creating a phonon ($\delta\hat{a}^{\dagger}\hat{b}^{\dagger}$),
although these will not be important for our scheme. For notational
simplicity (and following convention), we will from now on replace
$\delta\hat{a}$ by $\hat{a}$.

In the solid-state, the largest values of $g_{0}$ have been reached
in optomechanical (OM) crystals \cite{Eichenfield2009,Safavi-Naeini2010APL,Gavartin2011PRL_OMC,2011_Chan_LaserCoolingNanomechOscillator,SafaviNaeini2014SnowCavity}.
These are free-standing photonic crystals, i.e. dielectric slabs with
an appropriate pattern of holes which creates complete optical and
mechanical band gaps. A local modification of the pattern of holes
generates a point defect where optical and mechanical modes can become
localized. The OM interaction between such a localized optical and
mechanical mode is described by Eq.~(\ref{eq:StandardOMHamiltonian}),
with $g_{0}$ on the order of around $1\,{\rm MHz}$.

Future optomechanical arrays \cite{2011_Heinrich_CollectiveDynamics,2011_Chang_SlowingAndStoppingLight_NJP,Xuereb2012,2013_Ludwig}
can be produced by fabricating a periodic array of such point defects,
in 1D or 2D. The localized modes on adjacent lattice sites will have
an evanescent overlap, leading to tunneling of photons and phonons
between sites $i$ and $j$ with rates $J_{ij}$ and $K_{ij}$, respectively
\cite{Safavi-Naeini2011,2011_Heinrich_CollectiveDynamics,2013_Ludwig}.
For photons such tunneling-induced transport between localized modes
has been demonstrated experimentally in photonic crystal coupled resonator
waveguide arrays \cite{2008_notomilargephotonarrays}.

Combining both the optomechanical interaction at each site $j$ as
well as the tunneling between sites, the generic optomechanical array
Hamiltonian \cite{2013_Ludwig,Chen2014} reads

\begin{equation}
\hat{H}/\hbar=\sum_{j}\Omega\hat{b}_{j}^{\dagger}\hat{b}_{j}-\Delta\hat{a}_{j}^{\dagger}\hat{a}_{j}-\left(g_{j}\hat{a}_{j}^{\dagger}\hat{b}_{j}+h.c.\right)+\hat{H}_{{\rm hop}}\,.\label{eq:HamiltonianRWA}
\end{equation}
The annihilation operators of photons and phonons are denoted by $\hat{a}_{j}$
and $\hat{b}_{j}$, where the site index $j=(j_{1},j_{2},s)$ will
include a sub-lattice label $s$ for a non-Bravais lattice. The $\hat{a}_{j}$
are already displaced by the classical steady-state light amplitude
$\alpha_{j}$, set by the laser amplitude (as explained above), and
the $\hat{b}_{j}$ are displaced by the static mechanical displacement
$\beta_{j}$, set by the constant radiation force, see Appendix \ref{appendix:input-output}.
The term $\hat{H}_{{\rm hop}}/\hbar=-\sum_{i,j}J_{ij}\hat{a}_{i}^{\dagger}\hat{a}_{j}-\sum_{i,j}K_{ij}\hat{b}_{i}^{\dagger}\hat{b}_{j}$
incorporates the hopping of photons and phonons between different
sites, and $\Delta$ is the laser detuning from the optical resonance,
$\Delta\equiv\omega_{L}-\omega_{{\rm cav}}$, as we have switched
to a frame rotating at the laser frequency. To be clear, we note that
the detuning $\Delta$ defined here already includes a static effective
shift of the optical resonance due to the mechanical displacement
$\beta_{j}$, which depends on the laser intensity and is found from
the self-consistent classical solution. This is already known well
for the standard optomechanical system \cite{Aspelmeyer2013RMPArxiv}.

The optomechanical interaction displayed in Eq.~(\ref{eq:HamiltonianRWA})
converts phonons into photons propagating inside the array (and vice
versa). The strength of these processes is described by a laser-tunable
coupling constant $g_{j}=g_{0}\alpha_{j}$ that is parametrically
enhanced by the light amplitude, where $\left|\alpha_{j}\right|^{2}$
would be the steady-state photon number in mode $j$. The amplitude
$\alpha_{j}$ will depend on the site $j$ for the case of an inhomogeneous
driving field, to be considered below. For the sake of simplicity,
we have omitted the Stokes transitions of the type $\hat{a}_{j}^{\dagger}\hat{b}_{j}^{\dagger}$,
where photon-phonon pairs are emitted (or annihilated). Stokes processes
are strongly suppressed in the parameter regime that will turn out
to be suitable for the topologically nontrivial phase (where $\Delta$
will be negative, corresponding to a ``red-detuned'' laser drive),
so we neglect them at first. Since the anti-Stokes processes considered
here conserve the total excitation number, Hamiltonian Eq.~(\ref{eq:HamiltonianRWA})
is equivalent to a single-particle Hamiltonian, and we will be able
to use the standard classification of topological phases \cite{Hasan2010RMP}.

\subsection{A Chern Insulator implemented in an Optomechanical Array}

In multi-mode OM systems, the optical backaction can be used to engineer
the effective mechanical interaction, which has been suggested to
pave the way to phononic quantum information processing (e.g. \cite{Schmidt2012,Habraken2012}).
In this context, the phonon hopping amplitudes are modified by the
new pathways which are opened by the OM interaction. A phononic excitation
can be virtually converted into a photon on site $i$, hop to site
$j$, and be converted back into a phonon on site $j$. From standard
perturbation theory, the probability amplitude associated with this
pathway is $J_{ij}g_{i}g_{j}^{*}/(\Omega+\Delta)^{2}$. Hence, a pattern
of phases in the optomechanical coupling $g_{j}$ can lead to a synthetic
gauge field for phonons in the form of effective hopping rates $K_{ij}^{({\rm eff)}}=K_{ij}+K_{ij}^{({\rm opt)}}$
that contain an optically-induced component with complex phases \cite{Habraken2012}.

Inspired by previous studies which have indicated that a staggered
magnetic field for particles on a Kagome lattice yields topologically
nontrivial phases \cite{Ohgushi2000,Green2010,PhysRevLett.104.066403,Koch2011},
our investigations focus on a Kagome optomechanical array. We choose
this geometry since it can be naturally implemented in 2D optomechanical
crystals based on the snowflake design \cite{2010_Safavi_2DSimultaneousBandgap,Safavi-Naeini2011},
which have been demonstrated in an experiment recently \cite{SafaviNaeini2014SnowCavity}.
The general approach described here is of course applicable to other
lattice geometries as well. The Kagome optomechanical array is sketched
in Fig.~\ref{fig:Setup}. The idea is to have hexagonal patches of
periodically arranged snowflake-shaped holes, with linear dislocation
defects forming at the edges between those patches. As has been shown
in the experiment \cite{SafaviNaeini2014SnowCavity}, a suitable modification
of the hole pattern inside the linear defect then creates a point-like
defect with localized modes. The nearest-neighbor coupling between
those modes will generate the connectivity of a Kagome lattice. Its
unit cell contains three sites ($s=A,B,C$) forming an equilateral
triangle (we set the side to $1$). Thus, the optomechanical band
structure will comprise altogether six bands, three of them photon-like
and three phonon-like.

Some general properties of the band structure can be deduced purely
from the symmetry of the Kagome lattice geometry, without assuming
anything about the range of the hopping or other details. The hopping
term $\hat{H}_{\mathbf{{\rm hop}}}$ maintains the time reversal ${\cal T}$,
the inversion symmetry ${\cal I}$ with respect to any corner of the
triangle, and the symmetry ${\cal C}_{3}$ (rotations by $n2\pi/3$
around the triangle center, $n\in\mathbb{Z}$). Then, in the absence
of the laser drive there is no optical or mechanical band gap: For
both the photons and the phonons the central band touches one of the
remaining bands (top or bottom) at the center of the Brillouin zone,
$\vec{\Gamma}=(0,0)$, and the other one at the symmetry points $\vec{K}=(2\pi/3,0)$
and $\vec{K'}=(\pi/3,\pi/\sqrt{3})$, where Dirac cones form. 
\begin{figure*}
\includegraphics[width=1.9\columnwidth]{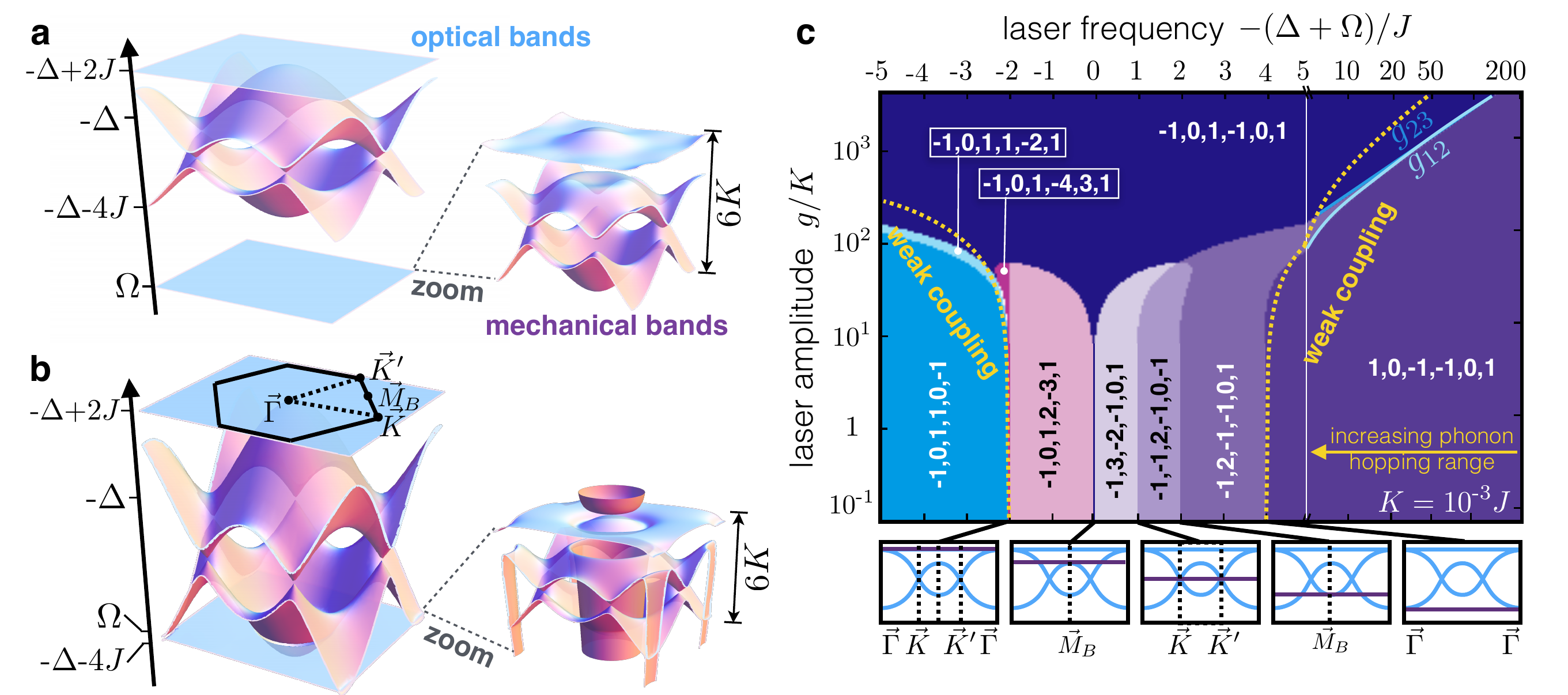}\protect\protect\caption{\label{fig:3Dbandstructureandphasediagram}(a) Band structure of a
Kagome optomechanical array, shown here in the case of well-separated
optical and mechanical bands (``weak-coupling limit''). The three
mechanical bands appear flat on the scale of the optical bands. A
zoom-in shows the resulting phonon insulator. (b) The ``strong-coupling''
limit where photon and phonon excitations mix. (c) Topological phase
diagram: The different topological phases are marked by the color
code and by the set of six Chern numbers, corresponding to the bands
ordered by frequency, uniquely identifying each phase. The schematic
band structures below the phase diagram indicate the symmetry points
where a pair of bands touch at the corresponding phase transition.
The OM coupling $g$ (set by the driving laser amplitude) is displayed
on a logarithmic scale. The scale of the laser frequency (expressed
via the detuning $\Delta$) is linear but switches to logarithmic
for large negative detunings, for clarity. The onset of weak coupling
is indicated by the line $g=0.1\Delta_{34}$, where $\Delta_{34}$
is the gap between optical and mechanical bands. The approximate analytical
expressions (see Appendix \ref{appendix:Chern}) for the boundaries
$g_{12}$ and $g_{23}$ of the intermediate topological phase introduced
by the long-range hopping are also shown. }
\end{figure*}

We now assume the lattice to be driven by a laser, with an optical
phase that depends on the site within the unit cell, leading to $g_{j}=ge^{i\varphi_{s}}$.
To retain the $\mathcal{C}_{3}$ symmetry, we choose a phase mismatch
of $2\pi/3$ between the sublattices, $\varphi_{B}-\varphi_{A}=\varphi_{C}-\varphi_{B}=\varphi_{A}-\varphi_{C}=2\pi/3$.
Physically, the driving laser phase pattern has to form a vortex around
each triangle center (Fig.~\ref{fig:Setup}). This can be achieved
via wave-front engineering of the impinging laser field. As opposed
to our recently proposed optomechanical generation of arbitrary synthetic
magnetic fields for photons \cite{SchmidtarXiv2013}, we emphasize
that here (i.e. for a Chern insulator of sound waves) only a single
laser frequency is needed and that the imprinted optical phase field
is periodic in space, greatly simplifying its generation. In fact,
one does not even need the versatility of a spatial light modulator
for this task. Superimposing three plane waves impinging on the sample
automatically creates the required pattern of optical phases, if they
are at $120$ degree angles with respect to each other within the
plane of the sample. This is illustrated in Fig.~\ref{fig:Setup}c
for the slightly simplified case of three interfering plane waves
of a scalar field, and we have confirmed that it also works when taking
into account the vector nature of the electromagnetic field.

\subsection{Band Structure and Topological Classification}

The resulting band structure in the presence of such a drive is shown
in Fig.~\ref{fig:3Dbandstructureandphasediagram}, and we will explain
it in more detail below. The site-dependent optomechanical interaction
breaks the time reversal symmetry, thereby opening large gaps between
the mechanical bands. As we will show, these gaps are topologically
nontrivial and lead to topologically protected sound waves propagating
at the edge of a finite system.

As discussed before, Hamiltonian (\ref{eq:HamiltonianRWA}) can be
translated from its second-quantized form into a single-particle version,
where the nature of the excitation (photon vs. phonon) is treated
as an internal state. Translational invariance permits us to rewrite
it in momentum space, using a plane wave ansatz: 
\begin{equation}
\hat{H}(\vec{k})/\hbar=\bar{\omega}-\delta\omega\hat{\sigma}_{z}/2-(\bar{t}+\delta t\hat{\sigma}_{z}/2)\hat{\tau}(\vec{k})-g(\hat{\mu}\hat{\sigma}_{x}+\hat{\nu}\hat{\sigma}_{y}).\label{eq:single-particle Hamiltonian}
\end{equation}
Here we have assumed that only nearest-neighbor sites are coupled,
although that (reasonable) approximation could be lifted without destroying
any of the essential physical properties discussed in the following,
see Appendix \ref{appendix:symmetryofkagome}. The binary degree of
freedom expressed by $\sigma_{z}=\pm1$ denotes photon (+1) vs. phonon
(-1) excitations, and $\hat{\sigma}_{x,y,z}$ are the Pauli matrices
in this subspace. Furthermore, we have introduced the parameters $\bar{\omega}=(\Omega-\Delta)/2$,
$\delta\omega=\Omega+\Delta$, $\bar{t}=(J+K)/2$, and $\delta t=(J-K)$.

The $3\times3$ matrices $\hat{\mu}$, $\hat{\nu}$ and $\hat{\tau}(\vec{k})$
in Eq.~(\ref{eq:single-particle Hamiltonian}) act on the sublattice
degree of freedom, referring to the three sites $s=A,B,C$ of the
unit cell. The Hermitean hopping matrix $\hat{\tau}(\vec{k})$ encodes
motion on the Kagome lattice, with $\tau_{AB}=1+e^{-i\vec{k}\vec{a}_{1}}$,
$\tau_{AC}=1+e^{i\vec{k}\vec{a}_{3}}$ and $\tau_{BC}=1+e^{-i\vec{k}\vec{a}_{2}}$,
where $\vec{a}_{1}=(-1,-\sqrt{3})$, $\vec{a}_{2}=(2,0)$, $\vec{a}_{3}=(-1,\sqrt{3})$
are the lattice basis vectors. At the symmetry points in the Brillouin
zone, the eigenbasis of the ${\cal C}_{3}$ rotations diagonalizes
$\hat{\tau}(\vec{k})$: the eigenvectors are the vortex $\left|\circlearrowleft\right\rangle \equiv\left|1,e^{i2\pi/3},e^{-i2\pi/3}\right\rangle /\sqrt{3}$,
the anti-vortex $\left|\circlearrowright\right\rangle \equiv\left|1,e^{-i2\pi/3},e^{i2\pi/3}\right\rangle /\sqrt{3}$,
and the vortex-free state $\left|\oslash\right\rangle \equiv\left|1,1,1\right\rangle /\sqrt{3}$.
The matrices $\hat{\mu}$, $\hat{\nu}$ describe the conversion between
photons and phonons. When the OM interaction converts a phonon into
an array photon, a driving photon is absorbed and its angular momentum
is transferred to the array photon. For example, a vortex-free phonon
$|M,\oslash\rangle$ is converted into a photon with a vortex, $|O,\circlearrowleft\rangle$
(M=''mechanical'', O=''optical''). The remaining allowed transitions
are $|M,\circlearrowleft\rangle\leftrightarrow|O,\circlearrowright\rangle$
and $|M,\circlearrowright\rangle\leftrightarrow|O,\oslash\rangle$.
All allowed transitions have matrix element $-g$, and this fully
specifies $\hat{\mu}$ and $\hat{\nu}$ in the $\mathcal{C}_{3}$
eigenbasis {[}see also Appendix \ref{appendix:symmetryofkagome}{]}.

The eigenfrequencies of Hamiltonian (\ref{eq:single-particle Hamiltonian})
form six photon-phonon polariton bands. The admixture between photon
and phonon bands is weak for all quasimomenta if the highest mechanical
and lowest optical bands are separated by a gap larger than the OM
coupling, $\Delta_{34}\equiv-\Delta-4J-\Omega+2K\gg g$, which we
will term the ``weak-coupling limit''. Then, the photons can be
adiabatically eliminated, arriving at an effective description for
the phonons which incorporates the optical backaction.

In the limit of both weak coupling ($\Delta_{34}\gg g$) and large
detuning, $\Delta_{34}\gg J$, the optically-induced effective phonon
hopping $K_{ij}^{({\rm eff})}$ will be restricted to nearest neighbors,
whence we arrive at the model investigated in \cite{Ohgushi2000,Green2010,Koch2011}.
A phonon hopping three times anticlockwise around a triangle, at each
step with probability amplitude $K_{ij}^{({\rm eff})}\approx K+e^{-i2\pi/3}Jg^{2}/(\Delta+\Omega)^{2}\equiv K^{({\rm eff})}e^{i\Phi/3}$,
picks up the phase (see Appendix \ref{appendix:effectivetightbinding})
\begin{equation}
\Phi=-\frac{3\pi}{2}+3\arctan\frac{2K(\Delta+\Omega)^{2}-Jg^{2}}{\sqrt{3}Jg^{2}}.\label{eq:syntheticflux}
\end{equation}
Keeping in mind that a vector potential $\vec{A}(\vec{r})$ imprints
the phase $q\hbar^{-1}\int_{\vec{r}_{i}}^{\vec{r_{j}}}\vec{A}(\vec{r})\cdot d\vec{r}$
on a particle with charge $q$ hopping on a lattice from $\vec{r}_{i}$
to $\vec{r}_{f}$, we interpret $\Phi$ as the (dimensionless) flux
of a synthetic gauge field piercing a triangle. Notice that there
is no net average magnetic field as the flux piercing a hexagon is
$-2\Phi$, see Fig~\ref{fig:Setup}. The flux $\Phi$ decreases monotonically
from $0$ to $-2\pi$ with the laser amplitude $g$. We emphasize
that in realistic implementations the photon hopping rate $J$ is
much larger than the phonon hopping rate $K$. It is precisely in
this limit that the construction adopted here works well (with interference
between direct phonon transport of amplitude $K$ and virtual transport
via the photonic route). Indeed, values for the phase all the way
down to $-2\pi$ can be\textbf{ }reached, staying well within the
weak-coupling limit where Eq.~(\ref{eq:syntheticflux}) has been
derived.

In the opposite, large-bandwidth limit, $J\gg\Delta_{34}$, only a
small quasimomentum region close to the $\vec{\Gamma}$ point contributes
to the optically induced mechanical hopping. Away from $\vec{\Gamma}$,
the OM interaction is suppressed, as the energetic distance between
the lowest optical band and the mechanical bands rapidly increases.
Thus, the effective mechanical hopping $K_{ij}^{({\rm eff})}$ is
long-range in this limit, and this will change the topological properties
to be discussed below. In general, the range of $K_{ij}^{({\rm eff})}$
is governed by the ratio $J/\Delta_{34}$ and can be tuned by changing
the gap $\Delta_{34}$ via the laser frequency.

Finally, going away from weak coupling, one can enter a regime where
photon and phonon bands cross and hybridize strongly. In the following,
we will discuss the topological properties of the optomechanical band
structure for all of these regimes.

For systems in the Quantum Hall state class A, which is realized here,
the topological state is uniquely identified by the bands' Chern invariants
(or TKNN invariant \cite{Thouless1982} after Thouless, Kohmoto, Nightingale
and den Nijs). They are defined as the integral over the Brillouin
zone of the Berry curvature of each band \cite{Thouless1982}: 
\begin{equation}
C_{l}=\frac{1}{2\pi}\int_{BZ}d^{2}k(\nabla_{\vec{k}}\times\vec{{\cal A}_{l}}(\vec{k}))\cdot\vec{e}_{z}\qquad l=1\dots6.\label{eq:Chernnumber}
\end{equation}
The Berry connection $\vec{{\cal A}}_{l}=i\langle\vec{k}_{l}|\nabla_{\vec{k}}|\vec{k}_{l}\rangle$
depends on the eigenstates $|\vec{k}_{l}\rangle$ of Hamiltonian (\ref{eq:single-particle Hamiltonian}),
describing hybrid excitations of photons and phonons. The full topological
phase diagram calculated numerically as a function of the laser parameters
is shown in Fig.~\ref{fig:3Dbandstructureandphasediagram}. Whenever
two (or more) bands touch, their Chern numbers may change, signaling
a topological phase transition.

We start our analysis from the regime of weak coupling, when optical
and vibrational bands are separated sufficiently (this regime is delimited
by the dotted yellow lines in Fig.~\ref{fig:3Dbandstructureandphasediagram}).
For concreteness, we focus on negative detunings $\Delta$, to the
right of the diagram. At the far right we are in the limit of nearest-neighbor
hopping. There the phonons are made to realize the Kagome Chern-insulator
model, with the flux $\Phi$ given by Eq.~(\ref{eq:syntheticflux}).
In this model, all three mechanical bands are separated by complete
band gaps. Both band gaps close simultaneously for special values
of the flux where time reversal symmetry is unbroken \cite{Koch2011}.
In our case, this happens when the laser is switched off, where $\Phi=0$,
and when it reaches a critical amplitude $g=g_{tp}\equiv(\Delta+\Omega)\sqrt{K/J}$,
where $\Phi=\pi$. The Chern numbers are $C_{1/3}=\pm{\rm sign[\sin(\Phi)]}$,
$C_{2}=0$, where the bands are ordered by increasing energy \cite{Ohgushi2000}.
Hence, a topologically non-trivial phase arises as soon as the driving
is switched on, and the system changes to a different topological
phase above the threshold $g_{tp}$. The photons also experience a
synthetic gauge field, whose flux can be obtained from Eq.~(\ref{eq:syntheticflux})
by exchanging $K$ and $J$ and changing the sign. This flux is therefore
small and has opposite direction. The photon band Chern numbers thus
turn out to be $C_{4/6}=\mp{\rm sign}[\Phi]$, $C_{5}=0$, without
any transition at $g_{tp}$\textbf{. }

When the photon and phonon bands come closer by changing the laser
detuning, an effective long-range hopping of phonons is induced optically,
as discussed above. Then, a new topological phase appears for intermediate
laser amplitudes, not predicted in the simple nearest-neighbor model.
The reason is that the mechanical band gaps do no longer close simultaneously
but at two different critical couplings $g_{12}$ and $g_{23}$ (Fig.~\ref{fig:3Dbandstructureandphasediagram}).
In the previously discussed limit of short range hopping, $J/\Delta_{34}\to0$,
these would again coalesce to become $g_{{\rm tp}}$. The Chern numbers
for long-range hopping can be computed analytically (see Appendix
\ref{appendix:Chern}). 
\begin{figure*}
\includegraphics[width=1.9\columnwidth]{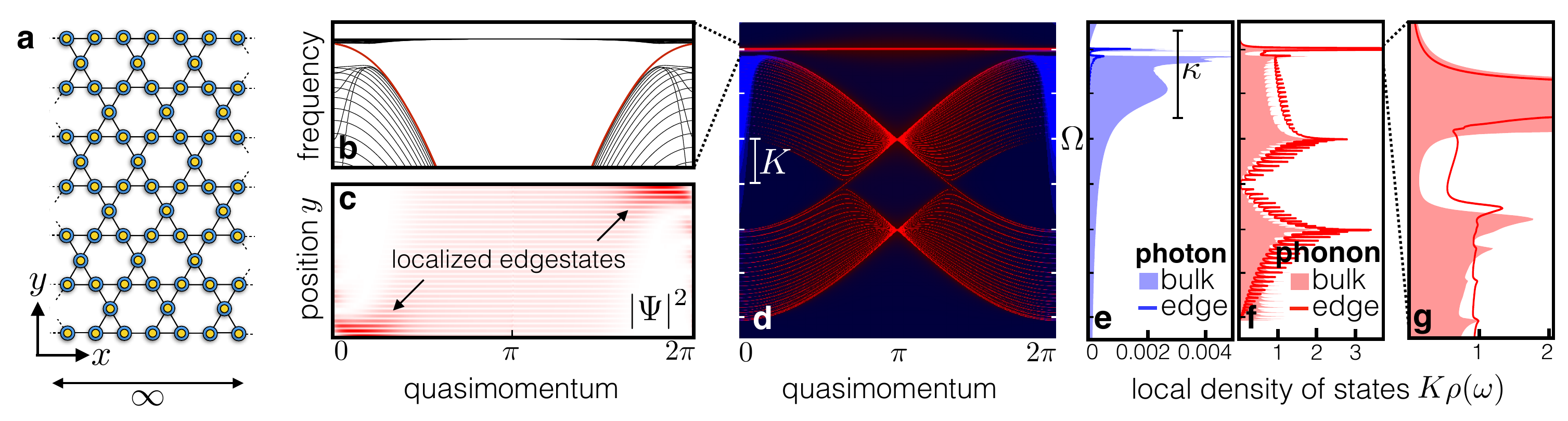}\protect\protect\caption{\label{fig:stripe}Edge states in a kagome optomechanical array. (d)
Band structure in the center of a finite-width strip, whose geometry
is shown in (a), as a function of the wavenumber along the longitudinal
direction of the strip. Blue/red indicates large photonic/phononic
components. Optical and mechanical dissipation, as well as the Stokes
interaction, have all been included (see Appendix \ref{appendix:Green}).
The indicated band gap is of topological nature. (b) Zoom-in of the
strip's band structure (here for clarity without dissipation). The
\foreignlanguage{american}{dispersion} of the edge states is highlighted.
The corresponding phonon probability density as a function of position
across the strip is shown in (c), demonstrating localization at the
edges for quasimomenta where the frequency lies in the bulk bandgap.
The photonic component (not shown) is small. (e) and (f) show the
local density of states for photons and phonons, respectively, both
in the bulk and at the edge of such a strip (of 30 unit cells width).
The band gap is much smaller than the photon decay rate $\kappa.$(g)
Zoom-in of the phonon local density of states. The parameters are
$\Omega=0.1J$, $\Delta=-4.02J$, $K=0.005J$, $g=0.007J$, $\kappa=0.01J$
and $\gamma=8\cdot10^{-5}J$.}
\end{figure*}

We now turn to the regime where the photon and phonon bands overlap
and interact strongly (see the center of the phase diagram Fig.~\ref{fig:3Dbandstructureandphasediagram}c).
There, the topological phases cannot be understood any more as induced
by an effective staggered synthetic gauge field for the phonons. They
give rise to a phase diagram that is unique for optomechanical arrays.
In this regime, a number of different phases appear. By inspecting
the limit of small coupling ($g\rightarrow0$), one notices that the
topological phase transitions occur whenever bands touch at the symmetry
points $\vec{\Gamma}$, $\vec{K},$ and $\vec{K}'$ or at the special
points $\vec{M}_{A}$, $\vec{M}_{B}$ and $\vec{M}_{C}$ (see the
sketch of the band structures at the phase transitions, bottom of
Fig.~\ref{fig:3Dbandstructureandphasediagram}c). This remains true
for arbitrary coupling, and can be explained as follows. Topological
phase transitions can happen whenever bands touch each other (instead
of repelling), which is possible if there are selection rules preventing
them from interacting. At the symmetry points $\vec{k}=\vec{\Gamma},\vec{K},\vec{K}'$
this is guaranteed by angular momentum conservation, whereas at $\vec{M}_{A}$
the optical and mechanical Kagome sublattice sites $A$ are decoupled
from the remaining sublattices $B$ and $C$ (likewise with $B$ at
$\vec{M}_{B}$ and $C$ at $\vec{M}_{C}$). The bands actually touch
simultaneously at $\vec{K},\vec{K}'$, due to inversion symmetry,
while rotational symmetry makes them touch simultaneously at $\vec{M}_{A}$,
$\vec{M}_{B}$ and $\vec{M}_{C}$. From these considerations, we can
predict the transitions to occur at the laser detunings $\Delta+\Omega\approx-4J,-2J,-J,0,2J$,
for small coupling $g$ and small mechanical hopping $K$. The resulting
set of Chern numbers for all the six bands is displayed in Fig.~\ref{fig:3Dbandstructureandphasediagram},
for each of the various topological phases.

\subsection{Chiral Edge State Transport}

A fundamental consequence of the topological nature of the optomechanical
band structure is the appearance of chiral edge states at the boundaries
of a finite size system. These excitations are topologically protected
against scattering if the bands are separated by a complete band gap.
They are thus very distinct from the type of edge states that are
produced in graphene-type systems with Dirac dispersion, which are
not robust against disorder and whose existence even depends on the
details of the boundary. The net number of such edge states (right-movers
minus left-movers) within a given band gap is directly determined
by the sum of the Chern numbers of all lower-lying bands. While in
the effective short-range Kagome model each pair of subsequent bands
are separated by such a gap, this is not generally true in the full
optomechanical model. Large gaps are desirable because they are more
robust against dissipation, disorder, and Stokes scattering, described
by additional terms $\hat{H}_{{\rm st}}=-\hbar\left(g_{j}\hat{a}_{j}^{\dagger}\hat{b}_{j}^{\dagger}+h.c.\right)$
in the Hamiltonian. However, topological band gaps cannot be larger
than the mechanical bandwidth $\sim K$ since they arise by the interplay
of intrinsic and optically induced hopping, see Appendix \ref{appendix:bandgapsize}.
For example, in the realistic regime where the optical bandwidth is
larger than the mechanical frequency the largest topological gap $\omega_{{\rm gap}}$
is given by $\omega_{{\rm gap}}\approx g\sqrt{2J/K}$ (we consider
a laser drive at the mechanical red sideband of the lowest frequency
optical mode), see Appendix \ref{appendix:bandgapsize}. In order
to show the feasibility of a topological optomechanical array, we
include all these aspects in the remainder of the paper, with a photon
decay rate $\kappa$ and a mechanical damping rate $\gamma$, see
Appendixes \ref{appendix:input-output} and \ref{appendix:Green}.
In Fig.~\ref{fig:stripe}, we show that for realistic parameters
the topological gaps are surprisingly resilient to dissipation. The
bulk band structure in Fig.~\ref{fig:stripe}d has a topological
gap between the second band (a hybrid photon-phonon band) and the
third band. The band gaps in the bulk photon and phonon local density
of states (LDOS), shown in Fig.~\ref{fig:stripe}e,f are weakly smeared
by dissipation, although the band gap is much smaller than the photon
decay rate $\kappa$. Such a robustness, which is related to the optomechanically
induced transparency phenomenon \cite{2010_Agarwal_EIT,2010_WeisKippenberg_OMIT,Safavi-Naeini2011SlowLight},
has been noticed in a different context already for a $1D$ optomechanical
array \cite{Chen2014}. It occurs because the excitations of the hybrid
band have strong phononic components at the band edge, making them
less sensitive to photon decay, see the color code in Fig.~\ref{fig:stripe}d.
In a strip of finite width (30 unit cells), one can observe that the
phononic wavefunctions form well-localized chiral edge states (Fig.~\ref{fig:stripe}b,c).
The residual bulk DOS inside the band gap (in Fig.~\ref{fig:stripe}f-g)
is produced when the mechanical dissipation smears the nearby large
peak in the DOS (the height of this peak is larger by a factor of
$\approx1600$ than the residual bulk DOS inside the bandgap). That
peak is formed by the localized excitations in the flat mechanical
band of the Kagome lattice. We have checked that the transport is
still chiral even in presence of such residual bulk DOS, see Appendix
\ref{sec:AppendixLossChannels}. Our analysis shows that the coupling
to the localized excitations causes injection losses but not backscattering.

\begin{figure}
\includegraphics[width=0.9\columnwidth]{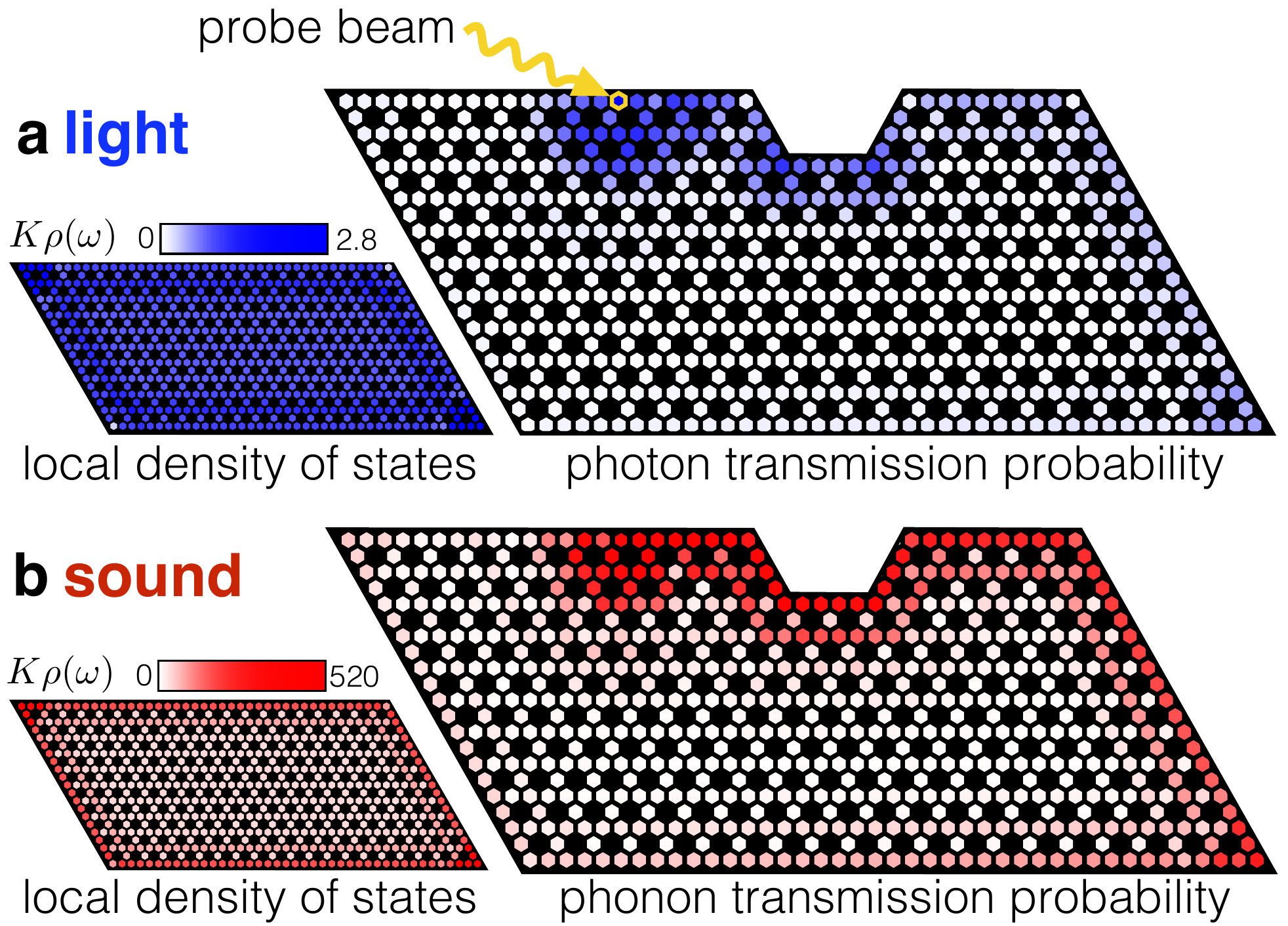}\protect\protect\caption{\label{fig:transport}Simulation of transport in a finite system.
A probe laser is injected locally at a site on the sample edge, at
a fixed frequency, launching phonons that are transported along the
chiral edge state. The larger figures depict the probability map of
finding a photon (a) or a phonon (b), respectively, demonstrating
transmission around an obstacle. Due to the optomechanical interaction,
the light intensity in (a), which could be imaged locally, represents
a faithful probe of the chiral phonon transport in (b). The probe
frequency lies in a bulk bandgap ($\omega_{{\rm L}}-\omega_{{\rm Probe}}=1.06J$)
that just permits mechanical edge states. The parameters correspond
to large detuning between optical and mechanical bands ($\Delta/J=-5.7$,
$\Omega/J=1$, $g/J=0.2$, $K/J=0.05$, $\kappa/J=0.1$ and $\gamma/J=0.002$).}
\end{figure}

Finally, we study transport in a finite-size array, for an experimentally
realistic setting that reveals the robustness against backscattering
by disorder. Additional robustness against mechanical dissipation
in the form of clamping losses can be provided by engineering 'phonon
shields', as demonstrated in \cite{2011_Chan_LaserCoolingNanomechOscillator}.
Since the gapless excitations at the sample edge are phononic in nature,
they could be excited by applying local oscillating stress. On the
other hand, in the current setting it is experimentally most straightforward
to shine light onto the sample edge. Even though the \emph{photon}
states are not localized at the edge (unlike the \emph{phonon} edge
modes), this simple approach works surprisingly well. Effectively,
the beat note between the weak, local probe laser and the strong,
global driving laser creates an oscillating radiation pressure force,
launching phonons. In Fig.~\ref{fig:transport}, we show a simulation
of topologically protected chiral sound waves excited locally by a
laser, traveling along the sample edge around an obstruction. In addition,
we have checked that also moderate random onsite disorder does not
affect the transport. Moreover, it turns out that spatially resolved
imaging of the light field emanating from the sample (Fig.~\ref{fig:transport}a)
can be used to map out the phonon edge state. This is because the
local vibrations will imprint sidebands on the strong drive laser,
and one of these sidebands appears at the probe laser frequency, which
then can be spectrally filtered and imaged.

In the strong-coupling regime discussed above, where photons and phonons
mix completely, one obtains chiral transport of photon-phonon polariton
excitations, which can also be excited and read out in the manner
discussed here.

The phonons will eventually decay, since the topological protection
prevents disorder-induced backscattering but not dissipation (the
same is true as well for all topological photon systems, for example).
The number of sites over which the phonons propagate along the chiral
edge state is given by their speed (the slope of the edge state dispersion)
divided by the overall mechanical decay rate (including both intrinsic
and optically induced dissipation). In the simplest case, the typical
scale of the propagation length is given by the ratio of the mechanical
hopping $K$ over the mechanical decay rate $\gamma$. For parameters
compatible with state-of-the-art devices the phonons can propagate
for about 100 sites before decaying, see Appendix \ref{sec:AppendixLossChannels}
for a more detailed analysis. This is completely sufficient for connecting
phonon reservoirs and other applications in phononics.

We now address briefly the most relevant issues for the experimental
realization of optomechanical Chern insulators. The most important
constraint in our model is the need to avoid a mechanical lasing instability
\cite{Aspelmeyer2013RMPArxiv} that may appear for larger bandwidths
$(J>\Omega/3)$ due to the Stokes terms. For laser frequencies below
the blue sideband of the lowest frequency optical mode, the instability
threshold is reached when the cooperativity equals (see Appendix \ref{appendix:lasing})
\begin{equation}
{\cal C}\equiv4\frac{g^{2}}{\kappa\gamma}=1+\left(\frac{\Delta+4J-\Omega-2K}{\kappa/2}\right)^{2}.
\end{equation}
The laser intensity (proportional to $g^{2}$) therefore has to remain
below this threshold, which was accounted for when selecting parameters
in our simulation results displayed here.

Regarding experimental parameters in general, in our figures we remain
compatible with those of the recent 2D snowflake crystal single-defect
experiment \cite{2010_Safavi_2DSimultaneousBandgap,SafaviNaeini2014SnowCavity},
where they report $\Omega\approx2\pi\times9{\rm GHz}$ and a single-photon
coupling strength of $g_{0}\approx2\pi\times250\,{\rm kHz}$. To obtain
the $g=g_{0}\sqrt{n_{{\rm phot}}}$ employed in Fig.~\ref{fig:stripe}
(where we assumed $J=10\Omega$ and $g=0.007J$) would require on
the order of $n_{{\rm phot}}\sim10^{6}$ circulating photons. Although
challenging, this should be doable, especially since any possible
increase of the phonon number due to unwanted heating (and finite
temperatures in general) does not affect measurements of the band
structure and transmission amplitudes, since the fluctuations do not
contribute to the average signal amplitudes.

In all future experiments on transport in optomechanical arrays, it
will be important to minimize disorder due to fabrication fluctuations,
and efforts to characterize and optimize this are only now starting.
In particular, post-fabrication processing techniques such as local
oxidation \cite{2011_Chen_SelectiveTuning} can be employed in the
future in order to drastically reduce the disorder by orders of magnitude.
In numerical simulations, we have seen that the topological effects
persist robustly up to disorder strengths of $2\%$ of $\Omega$ in
the mechanical on-site frequencies and of up to about $J$ in the
optical on-site frequencies (at $J=10\Omega$). More generally, we
have observed in our simulations that there is a wide latitude in
parameter combinations to obtain the effects discussed here. For example,
it may be more convenient experimentally to use larger photon hopping
rates $J$. Then, the instability is reached for smaller $g$, the
band gaps are smaller, and the edge states' penetration length is
larger. We checked that for $J=100\Omega$ and $g=5\times10^{-2}\Omega$
(and the other parameters in the same range as Fig.~\ref{fig:stripe}),
one can still\textbf{ }find a topological band gap. The corresponding
edge states are well localized on a strip of width $60$ unit cells.

\section{Conclusion}

Apart from its fundamental interest, chiral phonon transport, robust
against disorder, could be useful for many settings. Among them are
the transport of phonons between localized long-lived vibrational
modes (forming robust 'phononic networks') and the study of quantized
heat transport \cite{Schwaab2000} in an unconventional setting (with
a 'one-way' connection between heat reservoirs). The realization of
a phonon Chern insulator would thus also enable the observation of
new physical phenomena relevant to phononics. In addition, the mechanism
we have employed is conceptually distinct from anything that has been
considered for photons, to the best of our knowledge. In fact, the
optomechanical route towards Chern insulators has major advantages
over other proposals that have been put forward for photons and which
one might try to translate to phonons: The optomechanical concept
is more flexible than geometry-based approaches \cite{Hafezi2011,Hafezi2013},
since the properties can be tuned quickly in-situ, and in contrast
to settings based on local electrical modulation \cite{Fang2012},
it does not require local wiring of any kind (which is hard to scale
up).

The flexibility of the approach proposed here, where the pattern of
the laser field determines the band structure, could be exploited
to generate more general layouts in-situ, where arbitrarily shaped
regions of different topological phases are produced, studying the
transport through the edge states that form at their interfaces, possibly
arranged in interesting interferometer configurations. Moreover, the
time-dependent local control of the band structure could be the basis
for quench experiments on topological phases of light and sound. Finally,
if future improvements in the coupling $g_{0}$ between single photons
and phonons were to permit entering the strong single-photon coupling
regime (with $g_{0}\sim\kappa,\, g_{0}\sim\Omega)$, optomechanical
fractional Chern insulators could be realized, being governed by strong
quantum correlations.

\section*{Acknowledgements}

We acknowledge support by the ERC Starting Grant OPTOMECH, by the
DARPA project ORCHID, and by the European Marie-Curie ITN network
cQOM. We thank Aashish Clerk, Alexander Altland, and Oskar Painter
for discussions.

\section*{}

\appendix
%dummy comment inserted by tex2lyx to ensure that this paragraph is not empty

\section{Derivation of the linearized Hamiltonian for the Kagome optomechanical
array\label{appendix:input-output}}

We start from the standard input-output formalism for an array of
optomechanical cells (each consisting of a vibrational and an optical
mode) on a Kagome lattice, driven by a laser with uniform intensity
and a pattern of phases $\varphi_{j}$. As we intend to linearize
around the classical solution, we first write down the equations of
motion for the classical fields (the quantum fields averaged over
quantum and classical fluctuations) in a frame rotating with the drive:
\begin{eqnarray}
\dot{\beta}_{j} & = & (-i\Omega-\gamma/2)\beta_{j}+ig_{0}|\alpha_{j}|^{2}+iK\sum_{\langle l,j\rangle}\beta_{l},\nonumber \\
\dot{\alpha}_{j} & = & (i\Delta^{(0)}-\kappa/2)\alpha_{j}+i2g_{0}\alpha_{j}{\rm Re}\beta_{j}+iJ\sum_{\langle l,j\rangle}\alpha_{l}\nonumber \\
 &  & +\sqrt{\kappa}e^{i\varphi_{j}}|\alpha^{({\rm in)}}|.\label{eq:stationaryfields-1}
\end{eqnarray}
Here, $j=(n,m,s)$, $n,m\in Z$, $s=A,B,C$ and $\langle j,l\rangle$
indicates the sum over nearest neighbor sites. Moreover, $g_{0}$
is the shift of the optical frequencies due to a single phonon (more
precisely, a zero-point displacement), $K$ ($J$) is the phonon (photon)
hopping rate, and $\gamma$ ($\kappa$) is the phonon (photon) decay
rate. The laser detuning is $\Delta^{(0)}=\omega_{L}-\omega_{{\rm phot}}^{(0)}$,
and $|\alpha^{({\rm in)}}|$ is the absolute value of the driving
field. The phases $\varphi_{i}$ are independent of the unit cell
but they depend on the sublattice site: $\varphi_{B}-\varphi_{A}=\varphi_{C}-\varphi_{B}=\varphi_{A}-\varphi_{C}=2\pi/3$.
Then, the stationary solutions of (\ref{eq:stationaryfields-1}) are
given by $\alpha_{A}=e^{-i2\pi/3}\alpha_{B}=e^{i2\pi/3}\alpha_{C}$,
where $\alpha_{A}$ is a solution of the third order polynomial equation
\begin{equation}
\alpha_{A}=\frac{ie^{i\varphi_{A}}\sqrt{\kappa}\alpha^{({\rm in)}}}{\Delta^{(0)}+4J+2g_{0}^{2}|\alpha_{A}|^{2}/(\Omega-4K)+i\kappa/2}.\label{eq:classicalsolution}
\end{equation}
Without loss of generality, we can choose the phase of $\alpha^{({\rm in})}$
to fix $\alpha_{A}>0$ real-valued (this amounts to a gauge transformation).

We now linearize the quantum Langevin equations (input-output equations
of motion) around the classical solutions. We find (where $\hat{H}'=\hat{H}+\hat{H}_{{\rm st}}$
contains also the Stokes interaction terms): 
\begin{eqnarray}
 &  & \dot{\hat{b}}_{j}=i\hbar^{-1}[\hat{H'},\hat{b}_{j}]-\gamma\hat{b}_{j}/2+\sqrt{\gamma}\hat{b}_{j}^{({\rm in)}}\nonumber \\
 &  & =(-i\Omega-\gamma/2)\hat{b}_{j}+ig_{j}^{*}\hat{a}_{j}+ig_{j}\hat{a}_{j}^{\dagger}+iK\sum_{\langle l,j\rangle}\hat{b}_{l}+\sqrt{\gamma}\hat{b}_{j}^{({\rm in)}},\nonumber \\
 &  & \dot{\hat{a}}_{j}=i\hbar^{-1}[\hat{H'},\hat{a}_{j}]-\kappa\hat{a}_{j}/2+\sqrt{\kappa}\hat{a}_{j}^{({\rm in)}}\nonumber \\
 &  & =(i\Delta-\kappa/2)\hat{a}_{j}+ig_{j}(\hat{b}_{j}+\hat{b}_{j}^{\dagger})+iJ\sum_{\langle l,j\rangle}\hat{a}_{l}+\sqrt{\kappa}\hat{a}_{j}^{({\rm in})}\quad\label{eq:fullLangevin}
\end{eqnarray}
where $g_{A}=g_{0}\alpha_{A}=e^{-i2\pi/3}g_{B}=e^{i2\pi/3}g_{C}$,
and the detuning $\Delta$ includes a small shift of the optical resonances
due to the average mechanical displacement induced by the radiation
pressure: $\Delta=\Delta^{(0)}+2g_{0}^{2}|\alpha_{A}|^{2}/(\Omega-4K)$.
The input fields $\hat{b}_{j}^{({\rm in})}$ and $\hat{a}_{j}^{({\rm in)}}$
describe the vacuum (and possibly, thermal) fluctuations. The Hamiltonian
$\hat{H}$ is given in Eq.~(1) of the main text, and together with
the Stokes terms $\hat{H}_{{\rm st}}=-\hbar\left(g_{j}\hat{a}_{j}^{\dagger}\hat{b}_{j}^{\dagger}+h.c.\right)$,
it produces the right hand side of the Langevin equations displayed
here (except the fluctuation and decay terms, which stem from the
interaction with the vibrational and electromagnetic environment).

\section{Symmetries of the Kagome Lattice\label{appendix:symmetryofkagome}}

The topological effects discussed in the main text do not depend qualitatively
on the details of the hopping interactions (there for concreteness
we have assumed that only nearest neighbor sites are coupled) provided
that the underlying inversion symmetry (around a corner of the triangle
forming the unit cell) and the ${\cal C}_{3}$ rotational symmetry
of the Kagome lattice are retained. This applies in particular to
the topological phase diagram in Fig. 2 of the main text. In our model,
a topological phase transition occurs when two bands touch (instead
of repelling) as a result of a selection rule. This happens at the
symmetry points $\vec{\Gamma}$, $\vec{K}$ and $\vec{K}'$ where
only three transitions are allowed by the $C_{3}$ symmetry: $|M,\oslash\rangle\leftrightarrow|O,\circlearrowleft\rangle$,
$|M,\circlearrowleft\rangle\leftrightarrow|O,\circlearrowright\rangle$,
and $|M,\circlearrowright\rangle\leftrightarrow|O,\oslash\rangle$.
Moreover, two bands can touch at the special points $\vec{M}_{A}$,
$\vec{M}_{B}$ and $\vec{M}_{C}$ where the inversion symmetry ensures
that the optical and mechanical Kagome sublattices $A$, $B$, or
$C$, respectively, are decoupled from the remaining sublattices.

When these symmetry are broken the phase diagram becomes qualitatively
different. For instance, unequal mechanical and/or optical eigenfrequencies
on the different sublattices break the ${\cal C}_{3}$ symmetry. This
symmetry breaking has a twofold effect. First, the bands do not touch
anymore at the symmetry points $\vec{\Gamma}$, $\vec{K}$ and $\vec{K}'$.
This first effect does not change qualitatively the phase diagram
when a small perturbation breaks the symmetry. In this case, the bands
touch in a neighborhood of $\vec{\Gamma}$, $\vec{K}$ and $\vec{K}'$,
and the borders of the corresponding topological phase transitions
are only slightly deformed. Second, the bands do not touch simultaneously
at $\vec{M}_{A}$, $\vec{M}_{B}$ and $\vec{M}_{C}$. Then, the border
of the corresponding topological phase transitions split into three
lines and new topological phases appear. This second effect induces
a qualitative change of the topological phase diagram even when only
a small perturbation breaks the symmetry.

\section{Derivation of the effective tight-binding phonon Hamiltonian for
large detunings \label{appendix:effectivetightbinding}}

Our aim in this section is to integrate out the optical field and
derive the effective Hamiltonian for the phonons. Various ways exist
for doing this and here we choose to eliminate the optical fields
from the equations of motion. In this section, we consider the regime
of nearest-neighbor effective phonon hopping at the far right and
far left of the phase diagram in Fig.~2 of the main text. For concreteness,
we focus on the far right region in the diagram, $-\Delta-\Omega\gg J$.
Since we want to include also Stoke processes we start from the linearized
Hamiltonian, Eq.~(1) of the main text. Keeping in mind that the optical
backaction is filtered by the mechanical band, it is convenient to
divide $\hat{a}_{{\rm j}}$ into its sidebands, 
\begin{equation}
\hat{a}_{{\rm j}}\equiv e^{-i\Omega t}\hat{a}_{j}^{(red)}+e^{i\Omega t}\hat{a}_{j}^{(blue)}+\delta\hat{a}\label{eq:sidebands}
\end{equation}
When the mechanical bandwidth is small, i.e. when $6K^{(eff)}\ll\Omega$
(where $K^{({\rm eff)}}$ is calculated below), $a_{red}(t)$ and
$a_{blue}(t)$ are slowly varying functions (as is $\delta\hat{a}$,
describing the intrinsic optical fluctuations) and one can neglect
their time derivative in the Heisenberg equation $\dot{\hat{a}}_{j}=i\hbar^{-1}[\hat{H},\hat{a}_{j}]$.
We find 
\begin{eqnarray}
e^{-i\Omega t}\hat{a}_{j}^{(red)} & = & -\frac{g_{j}}{\Delta+\Omega}\hat{b}_{j}+\sum_{\langle j,l\rangle}\frac{Jg_{l}}{(\Delta+\Omega)^{2}}\hat{b}_{l}\nonumber \\
e^{i\Omega t}\hat{a}_{j}^{(blue)} & = & -\frac{g_{j}}{\Delta-\Omega}\hat{b}_{j}^{\dagger}+\sum_{\langle j,l\rangle}\frac{Jg_{l}}{(\Delta-\Omega)^{2}}\hat{b}_{l}^{\dagger}\label{eq:adiabaticfields}
\end{eqnarray}
We eliminate the photons by substituting Eqs.~(\ref{eq:sidebands}-\ref{eq:adiabaticfields})
in the Heisenberg equation $\dot{\hat{b}}_{j}=i\hbar^{-1}[\hat{H},\hat{b}_{j}]$
and arrive at $\dot{\hat{b}}_{j}=i\hbar^{-1}[\hat{H}_{{\rm eff}},\hat{b}_{j}]$
where 
\begin{eqnarray}
\hat{H}_{eff} & /\hbar\approx & \sum_{j}\Omega^{({\rm eff)}}\hat{b}_{j}^{\dagger}\hat{b}_{j}-\sum_{\langle j,l\rangle}K_{jl}^{({\rm eff)}}\hat{b}_{j}^{\dagger}\hat{b}_{l},\\
\Omega^{({\rm eff)}} & = & \Omega+\frac{g^{2}}{(\Delta+\Omega)}+\frac{g^{2}}{(\Delta-\Omega)},\\
K_{jl}^{({\rm eff)}} & = & K+J\frac{g_{j}^{*}g_{l}}{(\Delta+\Omega)^{2}}+J\frac{g_{j}g_{l}^{*}}{(\Delta-\Omega)^{2}}.\label{eq:effectivedampingwithStokes}
\end{eqnarray}
In deriving this, we have neglected the terms containing two creation/annihilation
operators (of the parametric oscillator type, $\hat{b}^{\dagger}\hat{b}^{\dagger}$
etc.), which is a good approximation for a small bandwidth $6K^{({\rm eff})}\ll\Omega^{({\rm eff})}$.
The third term in the right-hand side of Eq.~(\ref{eq:effectivedampingwithStokes})
describes the additional hopping amplitude induced by Stokes scattering
(going beyond the simpler approximation discussed in the main text,
where this term was neglected). The corresponding flux is 
\[
\Phi=-\frac{3\pi}{2}+3\arctan\frac{K(\Delta+\Omega)^{2}(\Delta-\Omega)^{2}-Jg^{2}(\Delta^{2}+\Omega^{2})}{-2\sqrt{3}Jg^{2}\Omega\Delta}
\]
The above result tends to the expression in Eq.~(3) of the main text
(which does not include Stokes scattering) for $\left|\Delta+\Omega\right|\ll\Omega$.
From this formula it is easy to prove that the small correction due
to the Stokes processes decreases the flux if $\Phi<-\pi$ but it
increases it if $\Phi>-\pi$. Since both mechanical band gaps reach
a maximum width at $\Phi=-\pi/2$ and $\Phi=-3\pi/2$, the Stokes
processes enlarge the gap in the broad parameter regime $-3\pi/2<\Phi<-\pi/2$.

\section{Calculation of the Chern numbers and the critical couplings in the
weak coupling regime\label{appendix:Chern}}

\subsection{Critical Couplings}

In the limit of a very large separation between optical and mechanical
bands, we have a model of phonons with effective nearest-neighbor
hopping on a Kagome lattice, and there is only one critical coupling
for a topological phase transition. When the separation is reduced,
longer-range hopping develops, and the first effect is that another
topological phase shows up. It is bounded by two critical couplings,
$g_{12}$ and $g_{23}$. These can be calculated by diagonalizing
the single-particle Hamiltonian Eq. (2) in the main text at the symmetry
points $\vec{\Gamma}$ and $\vec{K'}$ (the inversion symmetry ensures
that the second and third band will touch simultaneously at $\vec{K}$
and $\vec{K}'$). Due to rotational symmetry there are only three
allowed transitions at the symmetry points: $|M,\oslash\rangle\leftrightarrow|O,\circlearrowleft\rangle$,
$|M,\circlearrowleft\rangle\leftrightarrow|O,\circlearrowright\rangle$,
and $|M,\circlearrowright\rangle\leftrightarrow|O,\oslash\rangle$.
Hence, the Hamiltonian is block-diagonal with three $2\times2$ blocks
and can be very easily diagonalized for arbitrary $g$. However, this
leads to a nonlinear equation for the border of the phase transitions
$g_{12}$ and $g_{23}$. Instead, we restrict ourselves to the weak-coupling
regime (limit of large separation between optical and mechanical bands),
where it is possible to find simple analytical expressions for the
critical couplings and to calculate the Chern numbers analytically.

At the $\vec{\Gamma}$ point, the spectrum of the single-particle
Hamiltonian Eq. (2) of the main text is (up to leading order in $g$):
\begin{eqnarray*}
E_{O\circlearrowright} & = & -\Delta+2J+\frac{g^{2}}{-\Delta+2J-\Omega-E_{M\circlearrowleft}},\\
E_{M\circlearrowleft} & = & \Omega+2K+\frac{g^{2}}{\Delta-2J+\Omega},\\
E_{O\circlearrowleft} & = & -\Delta+2J+\frac{g^{2}}{-\Delta+2J-\Omega-E_{M\oslash}},\\
E_{M\oslash} & = & \Omega-4K+\frac{g^{2}}{\Delta-2J+\Omega},\\
E_{O\oslash} & = & -\Delta-4J,\qquad E_{M\circlearrowright}=\Omega+2K+\frac{g^{2}}{\Delta+4J+\Omega}.
\end{eqnarray*}
Here, we indicate with $E_{O\circlearrowright}$ the eigenvalue corresponding
to eigenvector $|\circlearrowright,O\rangle+\alpha|\circlearrowleft,M\rangle$
(with $\alpha\propto g$), and likewise for the other eigenvalues.
The above eigenvalues, ordered by increasing energy, are (for small
$g$): 
\begin{eqnarray*}
E_{1} & = & E_{M\oslash},\quad E_{2}=E_{M\circlearrowright},\quad E_{3}=E_{M\circlearrowleft},\\
E_{4} & = & E_{O\oslash},\quad E_{5}=E_{O\circlearrowleft},\quad E_{6}=E_{O\circlearrowright}.
\end{eqnarray*}
The coupling $g_{12}$ where the first and the second mechanical band
touch each other can be obtained from the condition $E_{M\oslash}=E_{M\circlearrowright}$,
yielding: 
\[
g_{12}=\left\{ 6K\left[(\Delta-2J+\Omega)^{-1}-(\Delta+4J+\Omega)^{-1}\right]^{-1}\right\} {}^{1/2}.
\]
Above this threshold, the first and second band exchange their eigenvectors,
\begin{eqnarray}
E_{1} & = & E_{M\circlearrowright},\quad E_{2}=E_{M\oslash},\quad E_{3}=E_{M\circlearrowleft},\nonumber \\
E_{4} & = & E_{O\oslash},\quad E_{5}=E_{O\circlearrowleft},\quad E_{6}=E_{O\circlearrowright}.\label{eq:bandstructuregammalongrangephase}
\end{eqnarray}

The same calculation at the $\vec{K}'$ point gives 
\begin{eqnarray*}
E_{O\circlearrowright} & = & -\Delta+2J,\qquad E_{M\circlearrowleft}=\Omega-K+\frac{g^{2}}{\Delta-2J+\Omega},\\
E_{O\circlearrowleft} & = & -\Delta-J+\frac{g^{2}}{-\Delta-J-\Omega-E_{M\oslash}},\\
E_{M\oslash} & = & \Omega-K+\frac{g^{2}}{\Delta+J+\Omega},\\
E_{O\oslash} & = & -\Delta-J+\frac{g^{2}}{-\Delta-J-E_{M\circlearrowright}},\\
E_{M\circlearrowright} & = & \Omega+2K+\frac{g^{2}}{\Delta+J+\Omega}.
\end{eqnarray*}
In this case, the eigenvalues ordered by increasing energy for small
$g$ are 
\begin{eqnarray}
E_{1} & = & E_{M\oslash},\quad E_{2}=E_{M\circlearrowleft},\quad E_{3}=E_{M\circlearrowright},\nonumber \\
E_{4} & = & E_{O\circlearrowleft},\quad E_{5}=E_{O\oslash},\quad E_{6}=E_{O\circlearrowright}.\label{eq:spectrumK'longrangephase}
\end{eqnarray}
The coupling $g_{23}$ where the second and third band touch each
other can be obtained from the condition $E_{M\circlearrowright}=E_{M\circlearrowleft}$,
yielding, 
\[
g_{23}=\left\{ 3K\left[(\Delta-2J+\Omega)^{-1}-(\Delta+J+\Omega)^{-1}\right]^{-1}\right\} {}^{1/2}.
\]
In the same way, at the $K$ point and for $g<g_{23}$ we have 
\begin{eqnarray}
E_{1} & = & E_{M\circlearrowright},\quad E_{2}=E_{M\oslash},\quad E_{3}=E_{M\circlearrowleft},\nonumber \\
E_{4} & = & E_{O\oslash},\quad E_{5}=E_{O\circlearrowright},\quad E_{6}=E_{O\circlearrowleft}.\label{eq:spectrumK'longrangephase-1}
\end{eqnarray}
Also at this point the second and third bands swap their eigenstates
at the critical coupling $g_{23}$.

\subsection{Chern numbers}

In the weak coupling regime, it is also possible to compute the Chern
numbers analytically. We will show this explicitly for the phase that
develops due to longer-range phonon hopping, i.e. the phase discussed
above between $g_{12}$ and $g_{23}$. We follow \cite{Kohmoto1985}.
Applying their general idea, we initially try to fix the gauge by
requiring $(\langle M,\circlearrowleft|+\langle O,\circlearrowright|)|\vec{k},l\rangle\in\mathbb{R}$,
where $|\vec{k},l\rangle$ is the eigenstate of band $l$ at $\vec{k}$.
If such a gauge were well defined over the whole Brillouin zone, the
Chern number would be $0$ {[}in Eq.~(4) of the main text, one integrates
the curl of a smooth function over a torus which gives zero from Stoke's
theorem{]}. However, there are \emph{obstructions} preventing us to
define a global gauge. At an obstruction the overlap $(\langle M,\circlearrowleft|+\langle O,\circlearrowright|)|\vec{k},l\rangle$
vanishes and the chosen gauge is ill defined. In its neighborhood,
i.e. a finite region within the Brillouin zone, one has to choose
a different gauge. In the new local gauge, the overlap $(\langle M,\circlearrowleft|+\langle O,\circlearrowright|)|\vec{k},l\rangle\equiv\rho(\vec{k})e^{-i\theta(\vec{k})}$
is a smooth function of $\vec{k}$ and its complex argument winds
an integer number of times $n$ on a path around the obstruction,
$n=(2\pi)^{-1}\oint\vec{\nabla}\theta(\vec{k})\cdot d\vec{k}$.\textbf{
}When calculating the Chern number, one picks up a contribution from
the boundary between the two regions of different gauge. The band
Chern number turns out to be the sum of the winding numbers for all
obstructions: $C_{l}=\sum_{i}n_{i}^{(l)}$. Such an analytical approach
is possible because, in the weak-coupling limit and for our particular
choice of gauge, obstructions form only at the symmetry points (this
does not hold in the strong coupling limit).

For concreteness, we focus on the second band and on the topological
phase introduced by the effective long-range hopping. As discussed
above, in this phase (corresponding to $g_{12}<g<g_{23}$), the second
band wavefunction is state $|M,\oslash\rangle$ (with a small admixture
to $|O,\circlearrowleft\rangle$) at the $\vec{\Gamma}$ point, state
$|M,\circlearrowleft\rangle$ (with a small admixture to $|O,\circlearrowright\rangle$)
at the $\vec{K}'$ point, and state $|M,\circlearrowright\rangle$
(with a small admixture to $|O,\oslash\rangle$) at the $\vec{K}$
point. Hence, for the second band, and for the global gauge defined
above, there are obstructions at $\vec{k}=\vec{\Gamma},\vec{K}$.

From the above discussion it is clear that in order to compute the
Chern number of the second band, it is sufficient to compute the overlap
$(\langle M,\circlearrowleft|+\langle O,\circlearrowright|)|\vec{k},2\rangle$
close to the symmetry points $\vec{\Gamma}$ and $\vec{K}$. We start
from $\vec{\Gamma}$. We decompose the Hamiltonian into $\hat{H}(\vec{\Gamma}+\delta\vec{k})=\hat{H}(\vec{\Gamma})-(\bar{t}+\delta t\hat{\sigma}_{z}/2)\delta\hat{\tau}_{\vec{\Gamma}}(\delta\vec{k})$
\begin{equation}
\delta\hat{\tau}_{\vec{\Gamma}}(\delta\vec{k})=i\begin{pmatrix}0 & -\delta\vec{k}\cdot\vec{a}_{1} & \delta\vec{k}\cdot\vec{a}_{3}\\
\delta\vec{k}\cdot\vec{a}_{1} & 0 & -\delta\vec{k}\cdot\vec{a}_{2}\\
-\delta\vec{k}\cdot\vec{a}_{3} & \delta\vec{k}\cdot\vec{a}_{2} & 0
\end{pmatrix},\label{eq:dtauGamma}
\end{equation}
where $\delta\vec{k}=\vec{k}-\vec{\Gamma}$. From Eqs. (\ref{eq:dtauGamma},\ref{eq:bandstructuregammalongrangephase})
we find, using standard perturbation theory in $\delta\vec{k}$: 
\[
(\langle O,\circlearrowright|+\langle M,\circlearrowleft|)|\hat{k},2\rangle\propto\langle\circlearrowleft|\delta\hat{\tau}_{\vec{\Gamma}}(\delta\vec{k})|\oslash\rangle\propto\delta k_{x}-i\delta k_{y}.
\]
Hence, the phase increases by $2\pi$ on a small path going anti-clockwise
around the obstruction: the winding number is $1$. In a neighborhood
of $\vec{K}$, we decompose the Hamiltonian into $\hat{H}(\vec{K}+\delta\vec{k})=\hat{H}(\vec{K})-(\bar{t}+\delta t\hat{\sigma}_{z}/2)\delta\hat{\tau}_{\vec{K}}(\delta\vec{k})$
\begin{eqnarray}
 &  & \delta\hat{\tau}_{\vec{K}}(\delta\vec{k})=\nonumber \\
 &  & i\left(\begin{array}{ccc}
0 & -e^{i2\pi/3}\delta\vec{k}\cdot\vec{a}_{1} & e^{-i2\pi/3}\delta\vec{k}\cdot\vec{a}_{3}\\
e^{-i2\pi/3}\delta\vec{k}\cdot\vec{a}_{1} & 0 & -e^{i2\pi/3}\delta\vec{k}\cdot\vec{a}_{2}\\
-e^{i2\pi/3}\delta\vec{k}\cdot\vec{a}_{3} & e^{-i2\pi/3}\delta\vec{k}\cdot\vec{a}_{2} & 0
\end{array}\right).\nonumber \\
\label{eq:dtauKprime}
\end{eqnarray}
From Eqs. (\ref{eq:spectrumK'longrangephase},\ref{eq:dtauKprime})
we find 
\[
(\langle O,\circlearrowright|+\langle M,\circlearrowleft|)|\vec{k},2\rangle\propto\langle\circlearrowright|\delta\hat{\tau}_{\vec{K}}(\delta\vec{k})|\circlearrowleft\rangle\propto\delta k_{x}-i\delta k_{y}.
\]
Notice that in this case the overlap comes from the optical part of
the wavefunction. From the above expression we see that the winding
number is again $1$. We can conclude that the Chern number for the
second band in the phase introduced by the long-range hopping (between
$g_{12}$ and $g_{23}$) is $2$. A similar calculation shows that
the first band has obstructions at $\vec{\Gamma}$ and $\vec{K}$
with winding number $-1$ and an obstruction at $\vec{K}'$ with winding
number $1$, whereas the third band has an obstruction with winding
number $-1$ at $\vec{K}'$. Hence, the Chern numbers for the mechanical
bands in the long range hopping phase are $[-1,2,-1]$. When the first
and second bands touch for $g=g_{12}$ at $\vec{\Gamma}$, the wavefunctions
change smoothly but they swap the bands. Hence, below $g_{12}$, also
the corresponding obstructions with their winding numbers are swapped
and we recover the result for small fluxes in the tight binding model:
$[1,0,-1]$. A similar argument shows that for the Chern numbers for
$g>g_{23}$ we recover the result for large fluxes in the tight binding
model: $[-1,0,1]$.

\section{Size of the band gaps\label{appendix:bandgapsize}}

\begin{figure*}
\includegraphics[width=1.9\columnwidth]{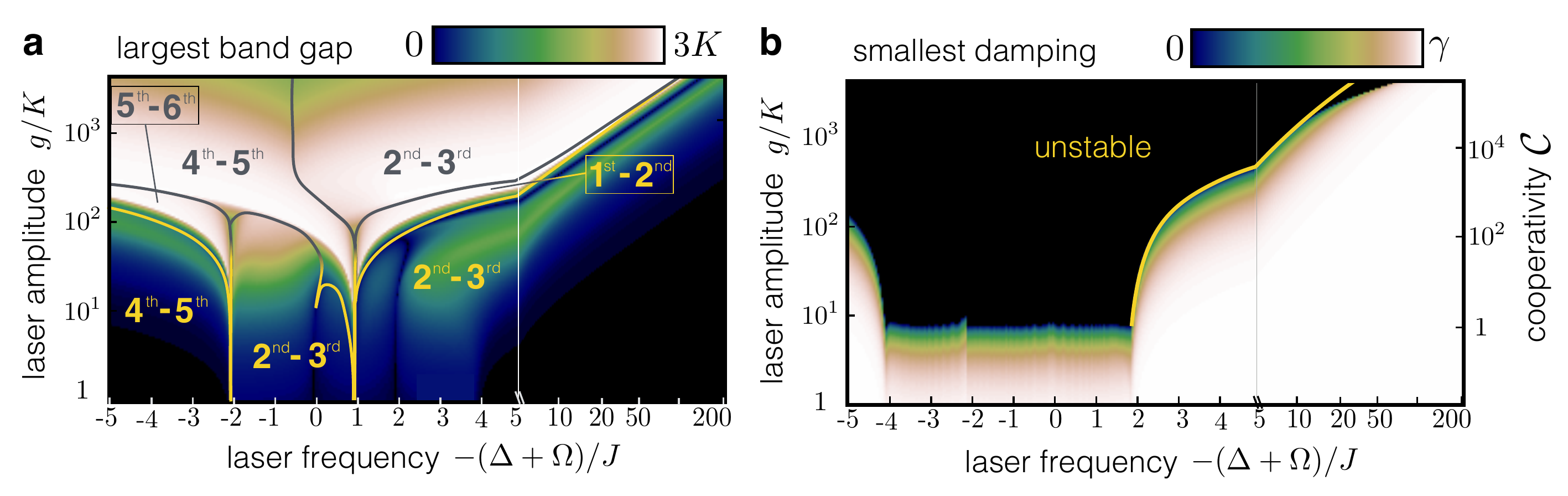}\protect\protect\caption{\label{fig:maximumgap}(a) Plot of the largest complete topologically
nontrivial band gap as a function of the laser parameters for $K/J=10^{-3}$.
The yellow lines divide the diagram in separate parameter regions.
The largest band gap lies between the subsequent bands indicated inside
each region. This diagram does not depend on $\Omega$. (b) Stability
diagram for $J/\Omega=1$, $\gamma=0.002J$, and $\kappa=0.1J$. It
shows the damping rate of the slowest relaxation process. The unstable
region where the Green function $\tilde{G}(\omega,l,j)$ has a pole
in the upper-half plane is marked in black. The mechanical lasing
threshold of Eq. (\ref{eq:lasingthreshold}) is plotted in yellow,
in its region of validity. Note that the onset of the mechanical lasing
instability (and that of another parametric instability visible in
the top right corner) restricts the region where the system is stable.
We emphasize that the stable region will include the whole parameter
range displayed in Fig. 2c of the main text for a sufficiently large
value of $\Omega$ (where $\Omega$ had not been specified for Fig.
2c, since that figure is independent of $\Omega$).}
\end{figure*}

The chiral excitations at the edge of the sample are more robust against
dissipation and disorder in the presence of large band gaps. In Fig.~\ref{fig:maximumgap},
we show the largest band gap as a function of the laser parameters.

Large band gaps of the order $\sim K$ are present for comparatively
large values of $g$, $g\gg K$. It is easy to understand this behavior:
for $K=0$ time reversal symmetry is not broken as one can eliminate
the pattern of phases in the couplings $g_{i}$ by a gauge transformation
on the phonon fields. In that case, all Chern numbers turn out to
be zero and there is no topologically nontrivial band gap. In the
presence of a small $K$ complete band gaps open. Since $K$ is the
smallest frequency scale, the band gaps can be computed by perturbation
theory in $K$ and are of order $\lesssim K$. For example, in the
limit of effective tight-binding phonon hopping (optical bands well
separated from mechanical bands), large band gaps are reached for
$\Phi=3/2$, where the size of the two mechanical band gaps is $3K$.

It is also possible to estimate the band gap in the more promising
parameter regime where the optical bandwidth is large $J\gtrsim\Omega$.
In this case, it is advantageous to choose the laser frequency such
that an optical band gets close but does not cross the mechanical
band, $\Omega\gg-\Delta-4J-\Omega>0$. In this regime, the modes in
the lower optical band and with quasimomentum $k\ll1$ interact most
strongly with the mechanics. In fact, all other optical modes are
far detuned due to the steep optical dispersion. Near $k=0$ (the
$\Gamma$ point), the low frequency optical modes have approximately
zero quasi-angular momentum. For a sideband resolved system we can
do a rotating wave approximation (since the blue sideband of the low
frequency optical modes is detuned by $2\Omega$). Since, a photon
with zero-quasiangular momentum is converted into a phonon with unit
angular momentum (the additional quasi-momentum comes from the laser
drive), only such mechanical mode is coupled to the light. Moreover,
at the $\Gamma$ point the mechanical states with quasi-angular momentum
$\pm1$ are quasidegenerate. Thus in order to compute the band gap
formed close to the $\Gamma$ point by the optomechanical interaction
we can neglect the influence of the remaining modes and project Hamiltonian
(\ref{eq:single-particle Hamiltonian}) into these three levels. In
a frame rotating with frequency $\Omega+2K$, the three levels are
described by the $3\times3$ effective Hamiltonian 
\[
\hat{H}_{{\rm eff}}=\begin{pmatrix}\delta\omega^{(O)}(\vec{k}) & -g & 0\\
-g & \delta\omega^{(M)}(\vec{k}) & \mathcal{K}(\vec{k})\\
0 & \mathcal{K}{}^{*}(\vec{k}) & \delta\omega^{(M)}(\vec{k})
\end{pmatrix}
\]
where $\delta\omega_{{\rm }}^{(O)}=-\Delta-4J-\Omega-2K+2Jk^{2}$,
$\delta\omega^{(M)}=-Kk^{2}$ and $\mathcal{K}=K(k_{x}+ik_{y})^{2}$.
If $\delta\omega_{{\rm }}^{(O)}(k)>0$ the optical band do not cross
the mechanical bands but pushes down the clock-wise phonon mode creating
a band gap. For very small $k$ (of the order of $g/J$) the bands
might also form polaritons. As the detuning increases the optical
interaction becomes weaker and tends to close the gap. The minimal
splitting is reached when the optically induced interaction is of
the same order as the coupling $\mathcal{K}(k)$ between the mechanical
modes with opposite quasi-angular momentum. For $|\delta\omega_{{\rm }}^{(O)}(k)-\delta\omega^{(M)}(k)|\gg|\mathcal{K}(k)|,g$,
we can eliminate adiabatically the low frequency optical mode. The
effective Hamiltonian for the remaining (mechanical) levels reads
\[
\tilde{H}^{({\rm eff)}}=\begin{pmatrix}\omega^{(M)}(\vec{k})-\frac{g^{2}}{\omega_{{\rm }}^{(O)}(\vec{k})} & \mathcal{K}(\vec{k})\\
\mathcal{K}^{*}(\vec{k}) & \omega^{(M)}(\vec{k})
\end{pmatrix}
\]
Thus, the eigenfrequencies of the second and third phononic bands
are (in the original frame) 
\[
E_{2/3}=\Omega+2K-Kk^{2}-\frac{g^{2}}{2\omega_{{\rm }}^{(O)}(k)}\mp\sqrt{\frac{g^{4}}{4\omega_{{\rm }}^{(O)2}(k)}+K^{2}k^{4}},
\]
independent of the direction of the quasi-momentum. The gap $\omega_{gap}$
between these two bands is given by the minimum of $E_{3}-E_{2}$
over the quasimomentum $k$. For concreteness we consider the case
where the red sideband of the lowest frequency optical mode coincides
with the largest frequency mechanical mode, $-\Delta-4J=\Omega+2K$.
In this case, we find a simple expression for the minimal splitting,
$\omega_{gap}\approx g\sqrt{2K/J}$.

In the most general case, we have computed numerically the largest
band gap as a function of the laser parameters. For fixed laser amplitude,
the largest band gap size varies on a broad range as a function of
the laser frequency, see Fig.~(\ref{fig:maximumgap}). Notice that
the mechanical eigenfrequency $\Omega$ is not specified in Fig.~\ref{fig:maximumgap}a.
It has been implicitly assumed to be the largest frequency in the
problem while neglecting the Stokes scattering (which involves a rotating
wave approximation), whence the band gaps become independent of $\Omega$.
Hence, the full phase diagram shown in Fig. 2 of the main text can
be explored for an appropriately large value of $\Omega$. On the
other hand, the effect of Stokes scattering has to be carefully analyzed
for large bandwidths $J/\Omega\gg1$ or large couplings $g^{2}\gg\Omega\kappa$.
Below, we show that the interplay of dissipation and Stokes scattering
restricts the laser parameter range where the system is stable. In
particular, we will focus on the experimentally most relevant regime
of large optical bandwidth, $J\gg\Omega$, where a mechanical lasing
transition arises.

\section{Calculation of the density of states and transmission probabilities
for a finite system\label{appendix:Green}}

In Fig.~3 and 4, we show the local densities of states (LDOS) on
site $l$, $\rho_{O}(\omega,l)$ for photons and $\rho_{M}(\omega,l)$
for phonons, as well as the probabilities $T_{OO}(\omega,l,j)$ and
$T_{MO}(\omega,l,j)$ that a photon (``O'' for optical) injected
on site $j$ is transmitted to site $l$ as a photon or a phonon (``M''
for mechanical), respectively. They are directly related to the retarded
Green's function in frequency space, $\tilde{G}(\omega,l,j)=\int_{-\infty}^{\infty}dte^{i\omega t}G(t,l,j),$
where the different interesting components are $G_{OO}(t,i,j)=-i\Theta(t)\langle[\hat{a}_{i}(t),\hat{a}_{j}^{\dagger}(0)]\rangle$
{[}propagation of a photon{]}, $G_{MO}(t,i,j)=-i\Theta(t)\langle[\hat{b}_{i}(t),\hat{a}_{j}^{\dagger}(0)]\rangle$
{[}conversion of a photon to a phonon{]}, and $G_{MM}(t,i,j)=-i\Theta(t)\langle[\hat{b}_{i}(t),\hat{b}_{j}^{\dagger}(0)]\rangle$
{[}propagation of a phonon{]}.

In order to calculate $\tilde{G}(\omega,l,j)$ numerically in a finite
system with $N\times M$ unit cells (see Fig.~4 of the main text),
one organizes all the fields in a $12NM$-dimensional vector $\vec{\hat{c}}$
whose entries are $\hat{a}_{j}$, $\hat{a}_{j}^{\dagger}$, $\hat{b}_{j}$,
$\hat{b}_{j}^{\dagger}$ for all possible $3NM$ lattice sites. Then,
Eq. (\ref{eq:fullLangevin}) can be written in a compact form as $i\dot{\vec{\hat{c}}}=A\vec{\hat{c}}+\vec{\hat{\xi}}$
and the Green function is $\tilde{G}(\omega)=(\omega-A)^{-1}$. Notice
that, in a system with $N\times M$ complete unit cells, the top and
right edges have a zig-zag form. In order to effectively describe
a system with only straight edges we set the hopping rates from and
to the sites on the zig-zag edges to zero. The photon and phonon LDOS
are given by 
\begin{eqnarray*}
\rho_{O}(\omega,l) & = & -2{\rm Im}\tilde{G}_{OO}(\omega,l,l),\\
\rho_{M}(\omega,l) & = & -2{\rm Im}\tilde{G}_{MM}(\omega,l,l),
\end{eqnarray*}
respectively. Moreover, from the Kubo formula and the input-output
relations $\hat{a}_{j}^{(out)}=\hat{a}_{j}^{(in)}-\sqrt{\kappa}\hat{a}_{j}$
(and likewise for the phononic fields), we find 
\begin{eqnarray*}
T_{OO}(\omega,l,j) & = & |\delta_{lj}-i\kappa\tilde{G}_{OO}(\omega,l,j)|^{2},\\
T_{MO}(\omega,l,j) & = & \kappa\gamma|\tilde{G}_{MO}(\omega,l,j)|^{2}\\
T_{MM}(\omega,l,j) & = & \gamma^{2}\left|\tilde{G}_{MM}(\omega,l,j)\right|^{2}
\end{eqnarray*}
For a strip that is infinite in the longitudinal direction and of
finite width $M$ unit cells (in Fig.~3 of the main text), the quasimomentum
in the longitudinal direction is a conserved quantity. Hence, the
LDOS is most conveniently calculated by taking a partial Fourier transform
of the corresponding index $n$ in Eq.~(\ref{eq:fullLangevin}).
For a numerical evaluation one considers a finite length $N$ and
introduces periodic boundary conditions for $n$, $\hat{a}_{j}=N^{-1/2}\sum_{n}e^{ikn}\hat{a}_{kms}$
(and likewise for $\hat{b}_{kms}$). For $N$ large enough the finite
size effects due to the finite length are smeared out by dissipation.
For the strip, we organize the fields $\hat{a}_{kms}$, $\hat{a}_{-kms}^{\dagger}$,
$\hat{b}_{kms}$, $\hat{b}_{-kms}^{\dagger}$ in a $12M$-dimensional
vector $\vec{\hat{c}}_{k}$. Then, the Langevin equation reads $i\dot{\vec{\hat{c}}}_{k}=A_{k}\vec{\hat{c}}_{k}$
and the corresponding Green's function is $\tilde{G}(\omega,k)=(\omega-A_{k})^{-1}$.
We arrive at the photon and phonon LDOS: 
\begin{eqnarray*}
\rho_{O}(\omega,n,s) & = & -2N^{-1}{\rm Im}\sum_{k}\tilde{G}_{OO}(\omega,k;n,s;n,s),\\
\rho_{M}(\omega,n,s) & = & -2N^{-1}{\rm Im}\sum_{k}\tilde{G}_{MM}(\omega,k;n,s;n,s).
\end{eqnarray*}

\section{Edge state transport: Analysis of Loss}

\label{sec:AppendixLossChannels}In this appendix we give more details
regarding the photon and phonon transport in the optomechanical array.
Our goal in this appendix is to analyze the propagation length of
phonons in the chiral edge states. In addition, we want to discuss
the appearance of a small but finite bulk density of states even inside
the bandgap. We argue that the directionality of the transport is
maintained in spite of that effect.

\begin{figure*}
\includegraphics[width=1\textwidth]{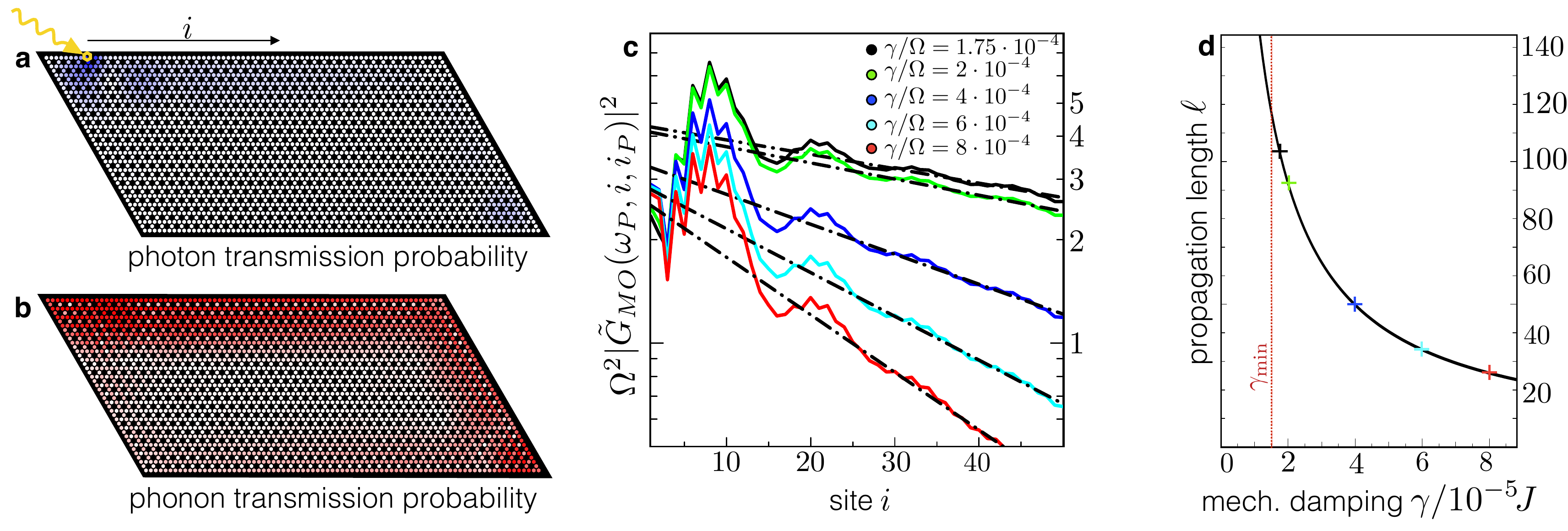}\protect\protect\caption{\label{fig:detailtransport}Photon and phonon transport in the regime
of large optical bandwidths. Panels (a,b) display the Green's functions
$\left|\tilde{G}_{OO}\right|^{2}$ and $\left|\tilde{G}_{MO}\right|^{2}$
for the propagation of photons and phonons, respectively, after injection
of a probe laser photon at a frequency $\omega$ inside the gap. Panel
(c) is a cut along the upper edge of picture (b), for several different
values of the mechanical damping $\gamma/\Omega=1.75\cdot10^{-4},\,2\cdot10^{-4},\,4\cdot10^{-4},\,6\cdot10^{-4},\,8\cdot10^{-4}$
{[}from top to bottom{]}, drawn on a log-scale to visualize the exponential
decay of phonon intensity. Panel (d) shows the propagation length
$\ell$ directly obtained by a fit of the data from panel (c) (symbols)
and by using the resulting data to fit the function $v/(\gamma+\gamma_{O})$
with the optically induced damping $\gamma_{O}$ as a fitting parameter
(solid line). The mechanical decay rate in (a,b) is $\gamma=4\cdot10^{-4}\Omega.$
All other parameters are those of Fig.~\ref{fig:stripe}, on a $20\times40$
array, at $\omega/J=0.10943$ (the middle of the bandgap).}
\end{figure*}

We focus on the most promising and realistic parameter regime where
the optical bandwidth is much larger than the mechanical eigenfrequency
(keeping the parameters of Fig.~\ref{fig:stripe} in the main text).
In order to obtain the phonon propagation length, we consider injection
at a particular site on the edge of a finite-size system (with a geometry
similar to Fig.~\ref{fig:transport}). In Fig.~\ref{fig:detailtransport},
we plot the decay of the phonon probability for different values of
the intrinsic mechanical decay rate $\gamma$. These values are compatible
with present-day experiments on optomechanical crystals, where even
higher mechanical Q factors ($10^{5}$ and more) are reached routinely
\cite{SafaviNaeini2014SnowCavity}. After some transient behavior
close to the injection point (where photons are converted into phonons),
the number of transmitted phonons decays exponentially with the propagation
distance. This allows us to extract the propagation length $\ell$
{[}see panel (d){]}. We expect $\ell$ to be given by the edge state
speed divided by the overall mechanical decay rate $\gamma_{{\rm total}}=\gamma+\gamma_{O}$
(the sum of the intrinsic and the optically induced mechanical decay
rates). By extracting $\ell$ from a fit of the numerical data we
find that indeed $\ell=v/\gamma_{{\rm total}}$, where the speed $v=2\partial\omega(k)/\partial k$
{[}with 2 sites along the edge per unit cell{]} is obtained from the
edge state dispersion in a strip geometry. The typical scale of $\ell$
is roughly given by the ratio of the phonon hopping $K$ over the
mechanical decay rate $\gamma$. This rule of thumb applies to the
broad parameter range where the optically induced mechanical hopping
and decay rates are at most of the order of their intrinsic counterpart.
From our fits we can also extract the optically induced damping which
turns out to be comparatively small {[}$\gamma_{O}/\Omega=4.2\cdot10^{-5}${]}.
We note in passing that $\gamma$ should be larger than a finite threshold
to avoid the mechanical lasing instability analyzed in Appendix \ref{appendix:lasing},
see Eq. (\ref{eq:lasingthreshold}). For the regime discussed here,
that would imply $\gamma>g^{2}\kappa/\Omega^{2}$, which we have ensured
to be true in the figures. This sets an upper limit on $\ell$ that
depends on the remaining parameters.

Next we comment on the transient behavior close to the injection point,
see Fig. \ref{fig:detailtransport}. The initial transient behavior
in the vicinity of the injection point (Fig. \ref{fig:detailtransport}c)
is partially due to photon-phonon conversion and the fact that we
chose to inject locally (at a single site). We now discuss an additional
effect during injection that is due to a residual contribution to
the bulk DOS inside the band gap, which, though small, is noticeable,
see also Fig. \ref{fig:stripe}g. We have found that this is due to
the broadening induced by mechanical dissipation. In particular, it
results from the tail of the nearby large and narrow DOS peak formed
by the localized excitations of the Kagome flat band. The peak is
so high that even a weak broadening can induce a non-negligible bulk
DOS inside the band gap. A phonon injected locally will tunnel not
only to the edge state but also (with a lower probability) to such
localized excitations (after which it will decay without moving far).
This is because local injection in principle can produce excitations
at any quasimomentum $k$, i.e. it also covers the full range of $k$
where the small dissipation-induced tail of the bulk band is present.
Apart from this influence on the injection process, the small finite
bulk DOS inside the gap can also have some influence during the propagation,
if there is disorder that is not smooth on the scale of the unit cell.
Then there can also be scattering with large momentum transfer that
will be able to scatter some fraction of the edge state excitations
into the dissipative tails of the localized bulk modes. We emphasize
though that the directionality of the transport is preserved in any
case (since phonons in the localized bulk modes do not contribute
to transport any more).

In addition, the injection losses could be reduced further by injecting
excitations in a momentum-resolved way, over a small interval of momenta.
Heisenberg's uncertainty relation then necessarily implies that they
cannot be injected at a single point but rather over an extended region;
e.g. by tunneling from a nearby phonon waveguide. In fact, to some
extent such a momentum-resolved injection even happens when exciting
the system optically, since the largest photon-phonon coupling occurs
in a limited range of quasimomenta (near $k=0$ for this parameter
regime).

\textcolor{black}{As for Fig.~\ref{fig:transport}, we note that
in comparison to Fig.~}\ref{fig:stripe}\textcolor{black}{{} we
had considered a smaller optical hopping $J$ (by one order of magnitude)
and a slightly larger optomechanical coupling $g$. This parameter
choice allows to display an edge state in a comparatively smaller
array, since a smaller $J$ implies shorter range optically induced
mechanical hopping and thus a smaller width of the edge states.}

\section{Mechanical lasing instability\label{appendix:lasing}}

Here, we show that the optical backaction can cause a mechanical lasing
instability for large enough couplings and optical bandwidths. There
is a phonon lasing instability if at least one mechanical mode, at
any point in the Brillouin zone, becomes unstable (negative damping
rate). Initially, we analyze the damping rates at the symmetry point
$\vec{\Gamma}$. There, it is convenient to write the OM interaction
in terms of the eigenstates of the ${\cal C}_{3}$ rotations with
quasimomentum $\vec{\Gamma}$: $\hat{b}_{\oslash}=({\cal N})^{-1/2}\sum_{l}\hat{b}_{l}$,
$\hat{b}_{\circlearrowleft}=({\cal N})^{-1/2}\sum_{l}e^{i\varphi_{l}}\hat{b}_{l}$,
$\hat{b}_{\circlearrowright}=({\cal N})^{-1/2}\sum_{l}e^{-i\varphi_{l}}\hat{b}_{l}$,
and likewise for $\hat{a}_{\oslash}$ $\hat{a}_{\circlearrowleft}$,
and $\hat{a}_{\circlearrowright}$ . Here and in the following, we
omit the quasimomentum label $\vec{\Gamma}$, ${\cal N}$ is the overall
number of sites forming the lattice, and the phases $\varphi_{l}$
only depend on the site A,B,C within the unit cell. The linearized
OM interaction reads 
\begin{eqnarray*}
H_{OM}(\vec{\Gamma}) & \approx & -g[\hat{a}_{\circlearrowright}^{\dagger}\hat{b}_{\circlearrowleft}+\hat{a}_{\oslash}^{\dagger}\hat{b}_{\circlearrowleft}^{\dagger}]-g[\hat{a}_{\oslash}^{\dagger}\hat{b}_{\circlearrowright}+\hat{a}_{\circlearrowright}^{\dagger}\hat{b}_{\circlearrowright}^{\dagger}]\\
 &  & -g[\hat{a}_{\circlearrowleft}^{\dagger}\hat{b}_{\oslash}+\hat{a}_{\circlearrowleft}^{\dagger}\hat{b}_{\oslash}^{\dagger}]+h.c.
\end{eqnarray*}
As it should be expected from quasi-angular-momentum conservation,
when a driving photon (which carries a vortex) emits a phonon with
a vortex $|\circlearrowleft,M\rangle$, it is simultaneously converted
into a vortex-free array photon (in the optical mode $|\oslash,O\rangle$,
with eigenfrequency $\omega_{{\rm phot}}-4J$) whereas when it absorbs
a phonon from the same mechanical mode it is converted into an array
photon with an anti-vortex (in the optical mode $|\circlearrowright,O\rangle$,
with eigenfrequency $\omega_{{\rm phot}}+2J$). This is a peculiar
situation, with different photon creation processes connected to phonon
absorption and emission, respectively. It can take place only because
the time-reversal symmetry is broken. Since the coupling strength
of both processes is the same (namely $-g$), we have anti-damping
of the mechanical mode $|\circlearrowleft,M\rangle$ if the blue sideband
frequency $\omega_{L}-\Omega$ is closer to the eigenfrequency of
$|\oslash,O\rangle$ than the red sideband frequency $\omega_{L}+\Omega$
is to the eigenfrequency of $|\circlearrowright,O\rangle$. In the
opposite situation we have damping.

There are two possible scenarios: The first scenario occurs for large
bandwidths, $J>\Omega/3$. Then, the blue sideband of the optical
mode $|\oslash,O\rangle$, located at $\omega_{{\rm phot}}-4J+\Omega$,
has lower frequency than the red sideband of $|\circlearrowright,O\rangle$,
located at $\omega_{{\rm phot}}-4J+\Omega$. In this case, the optical
backaction tends to amplify the mechanical mode $|\circlearrowleft,M\rangle$
when the driving is \emph{red detuned} (its frequency is below the
average eigenfrequency of the two optical modes, $-\Delta-J>0$).
Instead, the mechanical mode is damped by the optical backaction for
a \emph{blue detuned} drive ($-\Delta-J<0$). This behavior is completely
opposite to the standard scenario in optomechanics. A similar analysis
shows that the mechanical mode $|\circlearrowright,M\rangle$ shows
the opposite behavior. Hence, for \emph{any} choice of laser frequency,
either $|\circlearrowleft,M\rangle$ or $|\circlearrowright,M\rangle$
are antidamped (provided $J>\Omega/3$). The optically-induced antidamping
grows with increasing coupling and eventually overcomes the intrinsic
damping, thus generating a mechanical lasing transition at a critical
coupling.

The second scenario occurs for small bandwidths, $J<\Omega/3$. Then,
the blue sideband near $|\oslash,O\rangle$ has a higher frequency
than red sideband of $|\circlearrowright,O\rangle$. In this case,
the optical backaction damps the mechanical mode $|\circlearrowleft,M\rangle$
for a red detuned laser ($-\Delta-J>0$) and amplifies it for a blue
detuned drive ($-\Delta-J<0$). That is the standard behavior in optomechanical
systems. A similar analysis shows that the mechanical mode $|\circlearrowright,M\rangle$
displays the same behavior. Since, at $\vec{\Gamma}$ the spacing
between the optical eigenstates is largest, the same conclusion can
be drawn for any momentum. We can conclude that in the small bandwidth
case, $J<\Omega/3$, there is a mechanical lasing transition for a
\emph{blue detuned} drive but not for a \emph{red detuned} drive.
Notice that the region where no unwanted mechanical lasing transition
is present for small bandwidth $J$ includes the central part of the
phase diagram Fig. (2) (where a number of different topological phases
appear), as well as the ``tight-binding limit'' region on the right
part of the diagram.

It is possible to analytically compute the threshold of the mechanical
lasing transition for large bandwidths $J\gg\Omega$ and when the
driving frequency is below the blue sideband of the lowest frequency
optical mode $|\oslash,O\rangle$ (at the $\vec{\Gamma}$ point),
$-\Delta-4J>-\Omega-2K$. Since the other blue sidebands have larger
detuning, the lasing transition is determined by the backaction of
$|\oslash,O\rangle$ on the mechanical mode $|\circlearrowleft,M\rangle$.
In order to get simple formulas, we neglect the backaction by the
optical modes $|\circlearrowright,O\rangle$ and $|\circlearrowleft,O\rangle$.
This is a good approximation as these modes are far detuned for a
large optical bandwidth. The Langevin equations for the modes $\hat{a}_{\oslash}$,
$\hat{b}_{\circlearrowright}$, and $\hat{b}_{\circlearrowleft}$
read 
\begin{eqnarray*}
\dot{\hat{a}}_{\oslash} & = & (i\Delta+i4J-\kappa/2)\hat{a}_{\oslash}+ig(\hat{b}_{\circlearrowright}+b_{\circlearrowleft}^{\dagger})+\sqrt{\kappa}\hat{a}_{\oslash}^{({\rm in)}},\\
\dot{\hat{b}}_{\circlearrowright} & = & (-i\Omega-i2K-\gamma/2)\hat{b}_{\circlearrowright}+ig\hat{a}_{\oslash}+\sqrt{\gamma}\hat{b}_{\circlearrowright}^{({\rm in)}},\\
\dot{\hat{b}}_{\circlearrowleft} & = & (-i\Omega-i2K-\gamma/2)\hat{b}_{\circlearrowleft}+ig\hat{a}_{\oslash}^{\dagger}+\sqrt{\gamma}\hat{b}_{\circlearrowleft}^{({\rm in)}}.
\end{eqnarray*}
As before, we divide $\hat{a}_{\oslash}$ into its blue and red sidebands
as well as its intrinsic quantum fluctuations (optical vacuum noise)
\begin{equation}
\hat{a}_{{\rm \oslash}}\equiv e^{-i\Omega t}\hat{a}^{(red)}+e^{i\Omega t}\hat{a}^{(blue)}+\delta\hat{a}\label{eq:sidebands-1-1}
\end{equation}
For a narrow mechanical bandwidth $\gamma\ll\Omega$, the operators
$\hat{a}^{(red)}$ and $\hat{a}^{(blue)}$ are slowly varying and
we can neglect their derivative in the first Langevin equation. Then,
we find 
\begin{eqnarray*}
\hat{a}_{\oslash} & = & \frac{ig}{\kappa/2-i(\Delta+4J+\Omega+2K)}\hat{b}_{\circlearrowright}\\
 &  & +\frac{ig}{\kappa/2-i(\Delta+4J-\Omega-2K)}\hat{b}_{\circlearrowleft}^{\dagger}+\delta\hat{a},
\end{eqnarray*}
where $\delta\hat{a}$ describes the vacuum noise. By substituting
in the second and third Langevin equations and performing a rotating
wave approximation, we find 
\begin{eqnarray*}
\dot{\hat{b}}_{\circlearrowright} & = & (-i\Omega_{\circlearrowright}^{({\rm eff})}-\gamma_{\circlearrowright}^{({\rm eff})}/2)\hat{b}_{\circlearrowright}+\hat{\eta}_{\circlearrowright},\\
\dot{\hat{b}}_{\circlearrowleft} & = & (-i\Omega_{\circlearrowleft}^{({\rm eff})}-\gamma_{\circlearrowleft}^{({\rm eff})}/2)\hat{b}_{\circlearrowleft}+\hat{\eta}_{\circlearrowleft}.
\end{eqnarray*}
Here $\hat{\eta}_{\circlearrowright/\circlearrowleft}$ contains the
intrinsic mechanical as well as the optically induced noise. The effective
eigenfrequencies $\Omega_{\circlearrowright/\circlearrowleft}^{({\rm eff})}$
and decay rates $\gamma_{\circlearrowright/\circlearrowleft}^{(eff)}$
are obtained as  
\begin{eqnarray*}
\Omega_{\circlearrowright}^{({\rm eff})} & = & \Omega+2K+\frac{g^{2}(\Delta+4J+\Omega+2K)}{(\kappa/2)^{2}+(\Delta+4J+\Omega+2K)^{2}},\\
\gamma_{\circlearrowright}^{(eff)} & = & \gamma+\frac{g^{2}\kappa}{(\kappa/2)^{2}+(\Delta+4J+\Omega+2K)^{2}},\\
\Omega_{\circlearrowleft}^{({\rm eff})} & = & \Omega+2K+\frac{g^{2}(\Delta+4J-\Omega-2K)}{(\kappa/2)^{2}+(\Delta+4J-\Omega-2K)^{2}},\\
\gamma_{\circlearrowleft}^{(eff)} & = & \gamma-\frac{g^{2}\kappa}{(\kappa/2)^{2}+(\Delta+4J-\Omega-2K)^{2}}.
\end{eqnarray*}
We reach the threshold of the mechanical lasing transition when the
smaller rate reaches zero:\textbf{ }$\gamma_{\circlearrowright}^{(eff)}=0$,
corresponding to a maximum tolerable cooperativity (before hitting
the instability) of 
\begin{equation}
{\cal C}=4\frac{g^{2}}{\kappa\gamma}=1+\left(\frac{\Delta+4J-\Omega-2K}{\kappa/2}\right)^{2}.\label{eq:lasingthreshold}
\end{equation}
Our formula holds for a laser driving frequency below the blue sideband
of the lowest frequency optical mode $|\oslash,O\rangle$, $-\Delta-4J>-\Omega-2K$.
The threshold cooperativity increases monotonically from ${\cal C}=1$
to infinity for decreasing laser frequency. Notice that ${\cal C}=1$
represents also the lower bound for the maximum tolerable cooperativity.
It is reached when the driving is close to the blue sideband of any
optical mode.

\section{Stability Diagram\label{appendix:stability}}

In the general case, each relaxation process towards the classical
solution Eq. (\ref{eq:classicalsolution}) is associated to a pole
of the Green function $\tilde{G}(\omega,l,j)$ lying in the lower-half
complex plane. The corresponding damping rate is given by twice the
distance of the pole from the real axis. A pole in the upper-half
plane is associated to an excitation with negative damping and signals
that solution Eq. (\ref{eq:classicalsolution}) is unstable. In Fig.
\ref{fig:maximumgap}, we plot the damping rate of the slowest relaxation
process as a function of the laser parameters for $J=\Omega$. The
unstable region where the Green function $\tilde{G}(\omega,l,j)$
has at least one pole in the upper-half plane is marked in black.
The analytical expression for the border of the mechanical lasing
instability Eq. (\ref{eq:lasingthreshold}) is plotted in yellow.
It is has been derived for laser frequencies below the blue sideband
of the lowest frequency optical mode (the right-hand side of the stability
diagram). In the central part of the diagram corresponding to the
strong coupling regime, the maximum tolerable cooperativity stay close
to its lower bound ${\cal C}=1$ because the driving frequency is
always close to the blue sideband of an optical state. In the left
hand part of the diagram, the driving frequency is larger than the
blue sideband of the largest frequency mode and the lasing thresold
starts to increase again. Notice that at the far right of the diagram
the solution become unstable for values of the cooperativity below
the threshold of the mechanical lasing instability Eq. (\ref{eq:lasingthreshold}).
In this regime, the instability is not induced by mechanical lasing
but by a parametric instability. In optomechanical arrays, parametric
instabilities can occur for $g^{2}\gtrsim\Omega\kappa$ \cite{SchmidtarXiv2013,Chen2014}.
They set a finite limit to the tolerable cooperativity also in systems
with a small bandwidth $ $ driven by a red detuned laser where no
mechanical lasing transition is present.

\end{document}